\documentclass[10pt,a4paper,final]{iopart}
\usepackage{iopams,setstack}  
\usepackage{graphicx,multirow}
\usepackage[breaklinks=true,colorlinks=true,linkcolor=blue,urlcolor=blue,citecolor=blue]{hyperref}
                   
\def\as{\alpha_s}

\newcommand\one{\leavevmode\hbox{\small1\normalsize\kern-.33em1}}

\newcommand{\gev}{\mathrm{GeV}}

\newcommand{\iab}{\mathrm{ab}^{-1}}

\def\slashchar#1{\setbox0=\hbox{$#1$}           % set a box for #1
   \dimen0=\wd0                                 % and get its size
   \setbox1=\hbox{/} \dimen1=\wd1               % get size of /
   \ifdim\dimen0>\dimen1                        % #1 is bigger
      \rlap{\hbox to \dimen0{\hfil/\hfil}}      % so center / in box
      #1                                        % and print #1
   \else                                        % / is bigger
      \rlap{\hbox to \dimen1{\hfil$#1$\hfil}}   % so center #1
      /                                         % and print /
   \fi}

\newcommand{\eg}{\textsl{e.g.}\;}

\newcommand{\be}{\begin{eqnarray*}}
\newcommand{\ee}{\end{eqnarray*}}

\newcommand{\bee}{\begin{eqnarray}}
\newcommand{\eee}{\end{eqnarray}}
\newcommand{\beeq}{\begin{equation}}
\newcommand{\eeeq}{\end{equation}}

\usepackage{booktabs,multirow,tabularx}
\usepackage{caption}
\usepackage{float}
\usepackage{placeins}
\usepackage{rotating}
\usepackage{lscape}
\renewcommand\arraystretch{1.3}

\def \lsim{\mathrel{\vcenter
     {\hbox{$<$}\nointerlineskip\hbox{$\sim$}}}}

\newcommand{\nn}{\nonumber}

\newcommand\sss{\scriptscriptstyle}

\newcommand{\tth}{t\bar{t}H}
\newcommand{\ttz}{t\bar{t}Z}
\newcommand\ytop{y_{t}}

\newcommand{\ben}{\begin{enumerate}}
\newcommand{\een}{\end{enumerate}}
\newcommand{\bit}{\begin{itemize}}
\newcommand{\eit}{\end{itemize}}
\newcommand{\bc}{\begin{center}}
\newcommand{\ec}{\end{center}}
\newcommand{\bq}{\begin{equation}}
\newcommand{\eq}{\end{equation}}
\newcommand{\bqa}{\begin{eqnarray}}
\newcommand{\eqa}{\end{eqnarray}}

\def\beq{\begin{equation}}
\def\beqn{\begin{eqnarray}}
\def\eeq{\end{equation}}
\def\eeqn{\end{eqnarray}}
\def\beal{\begin{align}}
\def\endal{\end{align}}
\renewcommand\as{\alpha_{\sss S}}

\begin{document}
\title{Measuring the Top Yukawa Coupling at 100 TeV}

\author{Michelangelo L. Mangano}
\address{CERN, PH-TH, 1211 Geneva 23, Switzerland}
\ead{michelangelo.mangano@cern.ch}

\author{Tilman Plehn}
\address{Institut f\"ur Theoretische Physik, Universit\"at Heidelberg, Germany}
\ead{plehn@uni-heidelberg.de}

\author{Peter Reimitz}
\address{Institut f\"ur Theoretische Physik, Universit\"at Heidelberg, Germany}
\ead{p.reimitz@thphys.uni-heidelberg.de}

\author{Torben Schell}
\address{Institut f\"ur Theoretische Physik, Universit\"at Heidelberg, Germany}
\ead{schell@thphys.uni-heidelberg.de}

\author{Hua-Sheng Shao}
\address{CERN, PH-TH, 1211 Geneva 23, Switzerland}
\ead{huasheng.shao@cern.ch}

\begin{abstract}
We propose a measurement of the top Yukawa coupling at a 100 TeV
hadron collider, based on boosted Higgs and top decays. We find that
the top Yukawa coupling can be measured to 1\%, with excellent handles
for reducing systematic and theoretical uncertainties, both from side
bands and from $t\bar{t}H/t\bar{t}Z$ ratios.
\end{abstract}

\submitto{\JPG}
\tableofcontents

\maketitle

%%%%%%%%%%%%%%%%%%%%%%%%%%%%%%%%%%%%%%%%%%%%%%%%%%%
\markboth{Measuring the Top Yukawa Coupling at 100 TeV}{}
%\markboth{Measuring the Top Yukawa coupling}
%%%%%%%%%%%%%%%%%%%%%%%%%%%%%%%%%%%%%%%%%%%%%%%%%%%
\section{Introduction}
\label{sec:intro}

After the discovery of a light and likely fundamental Higgs boson
during the LHC Run~I~\cite{higgs,discovery}, the test of the Standard
Model nature of this Higgs boson will be one of the key goals of the
upcoming LHC run(s). One of the most interesting parameters of the
Standard Model (SM) 
is the top Yukawa coupling $\ytop$. One reason is that, because
of its large size, it dominates the renormalization group evolution of
the Higgs potential to higher, more fundamental energy
scales~\cite{higgs_rg}. On the other hand, this coupling is one of the
hardest to directly determine at
colliders~\cite{higgs_fit,Brock:2014tja}, 
because
this requires a precise measurement of the $t\bar{t}H$ production
cross section. This cross section can in principle be measured at
hadron colliders~\cite{heptop_1,tth_lhc,tth_had} as well as at $e^+
e^-$ colliders~\cite{sfitter_future,tth_lep}. However, a suitable $e^+
e^-$ collider should at least have an energy of 500~GeV. If a future
$e^+ e^-$ Higgs factory should have lower energy, the precise measurement of
$\ytop$\  will have to be postponed to a future hadron
collider, such as the 100~TeV $pp$ collider under consideration 
at CERN~\cite{FCC-hh} and in China~\cite{SppC}.\bigskip

The global set of physics opportunities of such a 100~TeV collider is
being explored in many studies.  Obvious pillars of the
physics program will include the study of weakly interacting thermal
dark matter~\cite{DM100TeV}, the gauge sector at high
energies~\cite{WZ100TeV}, the complete understanding of the nature
of the electroweak phase transition~\cite{EWPT}, and shedding more
light on the hierarchy problem.  The picture will rapidly
evolve in the near future, also in view of the forthcoming results for
the search of new physics at the LHC, in the experiments dedicated to
the study of flavor and CP violating phenomena, and at the
astro/cosmo frontier.  Nevertheless, the continued study of Higgs
properties, pushing further the precision of LHC measurements,
exploring rare and forbidden decays, and unveiling the whole structure
of the electroweak symmetry-breaking sector~\cite{BSMhiggs}, will provide the
underlying framework for the whole program.

These goals and benchmarks are, already today, clearly defined,
allowing us to start assessing their feasibility.  For example, first
studies indicate that a SM Higgs self-coupling could be measured at
100~TeV with a precision of 5-10\%~\cite{Higgs_self}, for an
integrated luminosity of $30~\iab$, consistent with the current
expectations~\cite{FCClumi}.  Similar 100~TeV studies, for the Higgs
couplings that are already under investigation at the LHC, are still
missing.  The fact that already at the high-luminosity LHC (HL-LHC)
the couplings' extraction will be dominated by systematic and
theoretical uncertainties~\cite{hl_lhc}, makes it hard to produce
today reliable predictions.  One important exception, where statistics
may still be limited at the HL-LHC, is $t\bar{t}H$ production. This
measurement is also a key ingredient for the determination of the
Higgs self-coupling. \bigskip

%--------------------------------------------------
\begin{figure}[b!]
\centering
 \includegraphics[width=0.46\textwidth]{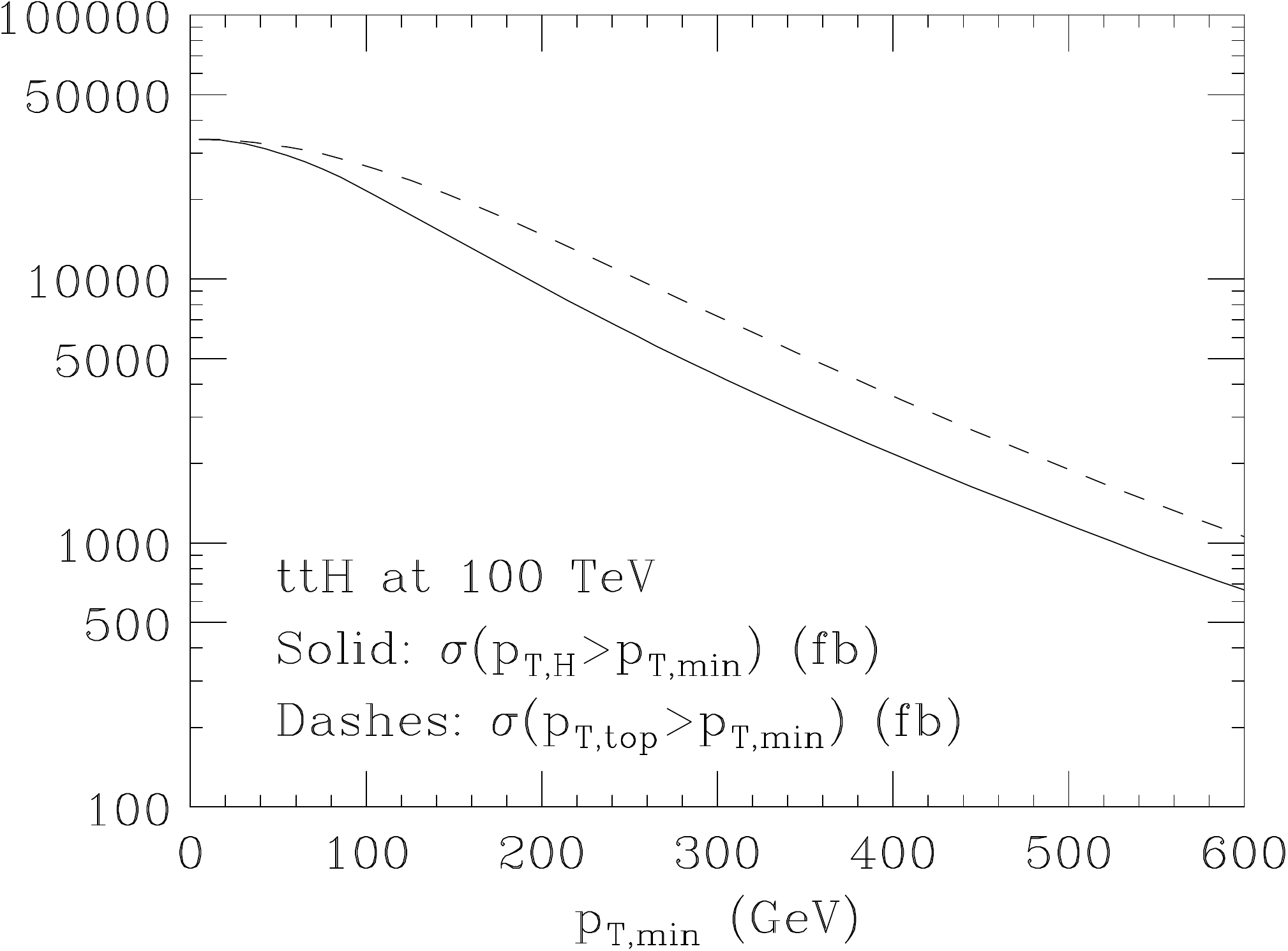}
 \hspace*{0.05\textwidth}
 \includegraphics[width=0.44\textwidth]{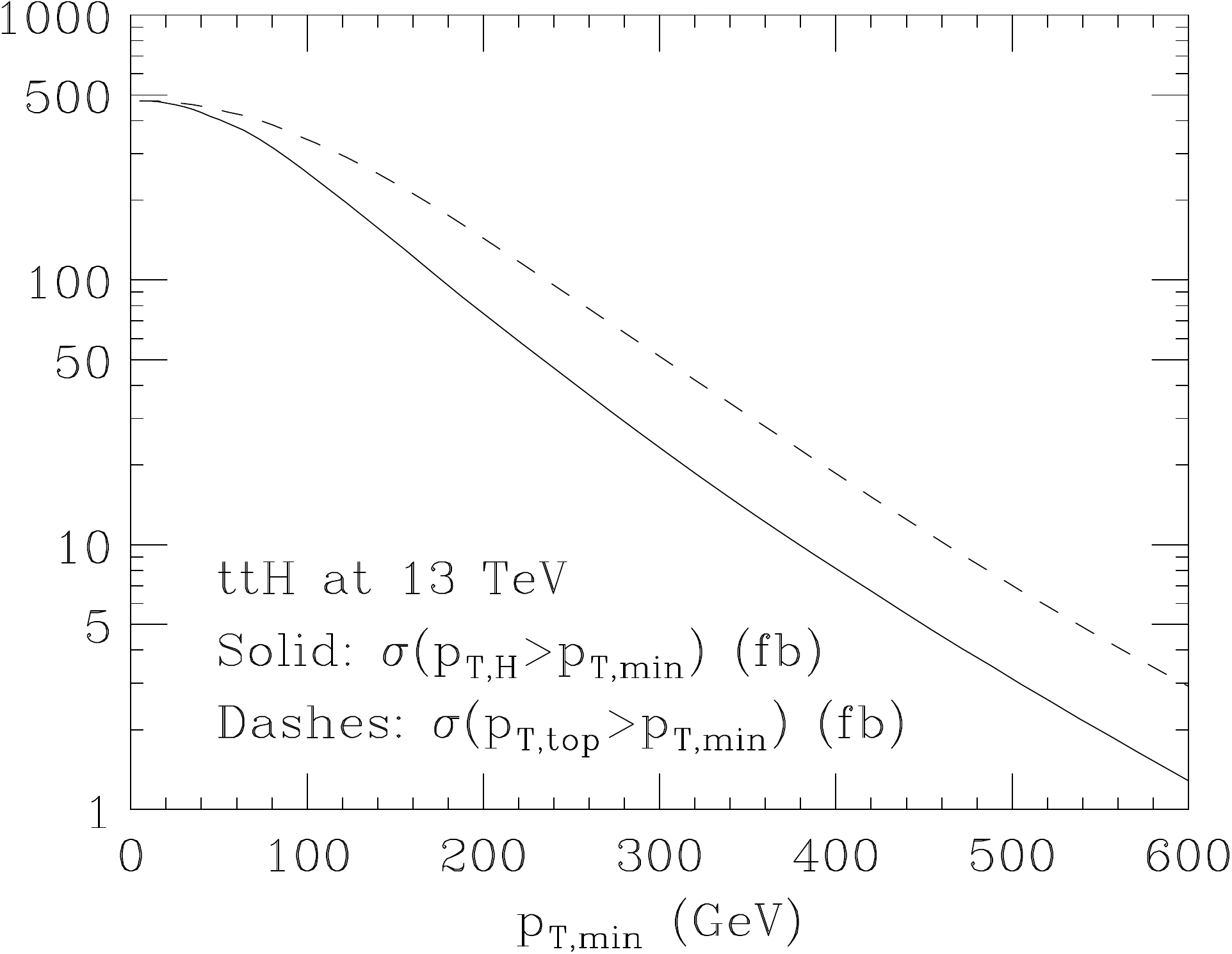}
\caption{Integrated transverse momentum distributions for the Higgs
  boson and top (anti-top) quark, in the $t\bar{t}H$
  process at a 100~TeV collider (left) and the 13~TeV LHC (right).}
\label{fig:pt}
\end{figure}
%--------------------------------------------------

In this paper we will show that a precision measurement of the top
Yukawa coupling $\ytop$\ should be added to the main physics
opportunities of a 100~TeV hadron collider. The crucial distinction
between this measurement at 100~TeV w.r.t. LHC energies is the potential to
fully exploit the features of boosted objects and jet
substructure~\cite{boosted}, thanks to a large-statistics sample of
highly boosted top and Higgs particles, as shown in Fig.~\ref{fig:pt}.
Our analysis will be based on the first \textsc{HEPTopTagger}
application to $t\bar{t}H$ production with a Higgs decay to
bottoms~\cite{heptop_1}.  There are three differences between the
original LHC analysis~\cite{heptop_1} and this 100~TeV
analysis: 

First, the statistically limited LHC analysis of boosted $t\bar{t}H$
production will benefit from the hugely increased statistics with a
100~TeV collider energy and an integrated luminosity of few tens of
$\iab$. For example, Fig.~\ref{fig:pt} shows that requesting
$p_{T,H}>500$~GeV gives a rate of ${\cal O}(1)$~pb, or 10M events with
$10~\iab$. This improved statistics also allows us to rely on a
well-measured and similarly peaked $t\bar{t}Z \to t\bar{t}\, b\bar{b}$
signal to reduce systematic and theoretical uncertainties.  In
particular, we will show in Section~\ref{sec:theory} that the cross
section ratio $\sigma(\tth)/\sigma(\ttz)$ is subject to very small
theoretical uncertainties, which, already today, are in the range of a
percent. This precision will certainly improve with future
calculations.

Second, in Fig.~\ref{fig:pt} we see that the typical transverse
momentum spectra of all particles are significantly harder, giving us
a larger relative fraction of events with $p_{T,t} > m_t$ and $p_{T,H}
> m_H$. The corresponding results with default taggers will be shown
in Section~\ref{sec:default}. 

Finally, the recent improvement in the
\textsc{HEPTopTagger2}~\cite{heptop_new} and in the BDRS Higgs
tagger~\cite{bdrs} will allow us to avoid background sculpting and to
increase the signal statistics. This last set of improvements will be
applied in Section~\ref{sec:update}. We will find that the ratio
between fiducial cross sections for the $\tth$ and $\ttz$ processes
can be measured with a percent-level statistical precision. Assuming
negligible beyond-the-SM contamination in the $t\bar{t}Z$ production
process, and in view of the theoretical systematics discussed in
Section~\ref{sec:theory}, this gives a measurement of the product
of $\ytop$\ times the $H \to b\bar{b}$ branching ratio, $B(H\to
b\bar{b})$, to 1\%. If the 100~TeV $pp$ collider will be preceded by
an $e^+e^-$ collider, $B(H\to b\bar{b})$ may be known to better than
1\%~\cite{Fujii:2015jha,Gomez-Ceballos:2013zzn,SppC,Linssen:2012hp}, 
providing a direct measurement of $\ytop$. If not, this result
will likely provide the most precise constraint on a combination of
Higgs couplings directly sensitive to $\ytop$.

%%%%%%%%%%%%%%%%%%%%%%%%%%%%%%%%%%%%%%%%%%%%%%%%%%%
\section{Theoretical systematics for the $\bf \tth/\ttz$ production rate}
\label{sec:theory}

It is well known that one of the key obstacles to exploiting the
immense statics available at hadron colliders for precision
measurements, is the intrinsic difficulty in performing accurate
absolute rate predictions. This difficulty arises from several
sources. On one side we have the complexity, and often the large size,
of higher-order contributions. At NLO one is often left with
uncertainties in the range of 10\% (although these can be much larger,
as in the case of Higgs or $b\bar{b}$ production, and more in general
for processes dominated by $gg$ initial states), uncertainties that
can be reduced to the few-percent level, but not always, only with the
inclusion of NNLO effects. On the other, there are uncertainties due
to the knowledge of initial-state parton distribution functions
(PDFs), which for $gg$ processes range from several percent, to
order-one factors in the case of very small or very large $x$
values. Furthermore, the modeling of the realistic final states,
including the description of hadronization and analysis cuts, which
are required for the comparison with experimental data, require an
additional layer of theoretical control, which very often cannot match
the available precision of fully inclusive parton-level
results. Finally, for specific processes, there are uncertainties due
to the knowledge of input parameters (e.g. the value of the top or
bottom mass, for processes involving these heavy quarks).

Over the past few years, we have witnessed nevertheless a
staggering progress in the theoretical precision, addressing all
aspects listed above~\cite{Dittmaier:2011ti}. 
A benchmark example is the recent completion~\cite{Anastasiou:2015ema} of
the NNNLO calculation for the inclusive Higgs production in the $gg\to
H$ channel, which, accompanied by the improved determination and
consistency of the gluon PDF luminosities~\cite{Rojo:2015acz}, 
has now reduced to about
$3\%_{\rm NNNLO}\oplus3\%_{\rm PDF}$ the current uncertainty on the total
production rate for this milestone process. A similar precision has
been achieved~\cite{Czakon:2013goa} 
in the case of the $t\bar{t}$ production rate, at
NNLO. In view of these examples, it is premature to establish today
what the theoretical systematics will be at the time a 100~TeV $pp$
collider will be operating. It is reasonable to anticipate that, also
thanks to the opportunities offered by the precise LHC measurements
for the validation of theoretical calculations and for the improvement
of the PDFs, within the next 10, 20 or 30 years all Higgs production
processes will be known with theoretical accuracy at the level of 1\%
or below.\bigskip

This notwithstanding, it is extremely useful to explore
observables that can help improving even further the precision, by
providing more robust confirmation of the systematics, and enabling
measurements where the experimental systematics can be reduced to
levels comparable to the theoretical ones. It is in this spirit that
we propose, for the study of this paper, the ratio of the $\tth$ to
$\ttz$ cross sections, performed in fiducial regions of acceptance
that make them suitable for a realistic experimental analysis. As we
shall discuss here, the theoretical understanding of these processes,
including NLO QCD~\cite{tth_nlo,tth_dec,ttz_nlo} and EW~\cite{tth_ew,ttz_ew}
 corrections, and including the current
knowledge of PDFs, allows already today to support an intrinsic
overall theoretical accuracy at the percent level. This precision will
certainly be consolidated, and further improved, by future
developments. Today, this allows to start probing the
experimental prospects of the 100~TeV collider, to put in perspective
the role of precision Higgs measurements at a such a facility, and to
provide useful performance benchmarks for the design of the future
detectors. In this Section we shall motivate such accuracy claim. What
will be learned, can also contribute to improve the expectations for
future runs of the LHC, by improving the predictions for the relative
size of the $\tth$ signal and its irreducible $\ttz$ background. 

%%%%%%%%%%%%%%%%%%%%%%%%%%%%%%%%%%%%%%%%%%%%%%%%%%%
\subsection{Total rates and ratios}

The main observation motivating the interest in the study of the
$\tth/\ttz$ ratio is the close analogy between the two processes. At
leading order (LO) they are both dominated by the $gg$ initial state,
with the $H$ or $Z$ bosons emitted off the top quark.
The $q\bar{q}$-initiated
processes, which at the 100 (13)~TeV amount for $\lsim 10\%$ ($\lsim
30\%$) of the total rates, only differ in the possibility to radiate
the $Z$ boson from the light-quark initial state. The difference
induced by this effect, as we shall see, is not large, and is greatly
reduced at 100~TeV. At NLO, renormalization, factorization and
cancellation of collinear and soft singularities will be highly
correlated between the two processes, since the relevant diagrams have
the same structure, due to the identity of the tree-level
diagrams. This justifies correlating, in the estimate of the
renormalization and factorization scale uncertainties, the scale
choices made for $\tth$ and $\ttz$. The uncertainties due to the mass
of the top quark are also obviously fully correlated between numerator
and denominator. Furthermore, due to the closeness in mass of the
Higgs and $Z$ bosons and the ensuing similar size of the values of $x$
probed by the two processes, and given that the choice of PDFs to be
used in numerator and denominator in the scan over PDF sets must be
synchronized, we expect a significant reduction in the PDF systematics
for the ratio. Finally, the similar production kinematics (although
not identical, as we shall show in the next Section), should guarantee
a further reduction in the modeling of the final-state structure, like
shower-induced higher-order corrections, underlying-event effects,
hadronization, etc.

%--------------------------------------------------
\begin{table}[b!]
 \begin{center}
 \renewcommand{\arraystretch}{1.3}
\begin{tabular}{r|ccc}
\hline
\rule{0pt}{3.6ex} & $\sigma(t\bar{t}H)$[pb] & $\sigma(t\bar{t}Z)$[pb] &
  $\frac{\displaystyle  \sigma(t\bar{t}H)}{\displaystyle \sigma(t\bar{t}Z)}$ \\ 
\hline
13 TeV	& $0.475^{+5.79\%+3.33\%}_{-9.04\%-3.08\%}$	&
$0.785^{+9.81\%+3.27\%}_{-11.2\%-3.12\%}$		&
$0.606^{+2.45\%+0.525\%}_{-3.66\%-0.319\%}$		\\
\hline
100 TeV	& $33.9^{+7.06\%+2.17\%}_{-8.29\%-2.18\%}$	&
$57.9^{+8.93\%+2.24\%}_{-9.46\%-2.43\%}$	&
$0.585^{+1.29\%+0.314\%}_{-2.02\%-0.147\%}$ \\ \hline	
\end{tabular}
\end{center}
\caption{Total cross sections $\sigma(\tth)$ and
  $\sigma(\ttz)$ and the ratios
  $\sigma(\tth)/\sigma(\ttz)$ with NLO QCD corrections at 13
  TeV and 100~TeV. Results are presented together with the
  renormalization/factorization scale and PDF$+\as$
  uncertainties.} 
\label{tab:scale}
\end{table}
%--------------------------------------------------

The above qualitative arguments are fully supported by the actual
calculations. All results are obtained using the {\sc MadGraph5\_aMC@NLO}
code~\cite{Alwall:2014hca}, 
which includes both NLO QCD and EW corrections.
The default parameter set used in this study is:

\begin{center}
 \renewcommand{\arraystretch}{1.0}
\begin{tabular}{cl|cl}\hline
Parameter & value & Parameter & value
\\\hline
$G_{\mu}$ & \texttt{1.1987498350461625} $\cdot$ \texttt{10${}^{-5}$} & $n_{lf}$ & \texttt{5}
\\
$m_{t}$ & \texttt{173.3}    & $y_{t}$ & \texttt{173.3}
\\
$m_{W}$ & \texttt{80.419}    & $m_{Z}$ & \texttt{91.188}
\\
$m_{H}$ & \texttt{125.0}    & $\alpha^{-1}$ & \texttt{128.930}
\\\hline
\end{tabular}
\end{center}
MSTW2008 NLO~\cite{Martin:2009iq} is the default PDF set 
and $\mu_R=\mu_F=\mu_0=\sum_{f\in {\rm final~states}}{m_{T,f}}/2$ is
the default for the central choice of renormalization and
factorization scales, where $m_{T,f}$ is the transverse mass of the final
particle $f$. This scale choice interpolates between the
dynamical scales that were shown in Ref.~\cite{tth_nlo} to minimize
the $p_T$ dependence of the NLO/LO ratios for the top and Higgs spectra.\bigskip

%--------------------------------------------------
\begin{table}[t]
 \begin{center}
 \renewcommand{\arraystretch}{1.3}
 \begin{tabular}{r|c|ccc}
\hline
\rule{0pt}{3.6ex} && $\sigma(t\bar{t}H)$[pb] & $\sigma(t\bar{t}Z)$[pb] &
   $\frac{\displaystyle \sigma(t\bar{t}H)}{\displaystyle \sigma(t\bar{t}Z)}$ \\ 
\hline
\multirow{3}{*}{13~TeV}
&   MSTW2008 & $0.475^{+5.79\%+2.02\%}_{-9.04\%-2.50\%}$ &
   $0.785^{+9.81\%+1.93\%}_{-11.2\%-2.39\%}$ &
   $0.606^{+2.45\%+0.216\%}_{-3.66\%-0.249\%}$ \\ 
&   CT10 & $0.450^{+5.70\%+6.00\%}_{-8.80\%-5.34\%}$ &
   $0.741^{+9.50\%+5.91\%}_{-10.9\%-5.29\%}$ &
   $0.607^{+2.34\%+0.672\%}_{-3.47\%-0.675\%}$ \\ 
&  NNPDF2.3 & $0.470^{+5.26\%+2.22\%}_{-8.58\%-2.22\%}$	&
  $0.771^{+8.97\%+2.16\%}_{-10.6\%-2.16\%}$	&
  $0.609^{+2.23\%+0.205\%}_{-3.41\%-0.205\%}$\\ \hline
\multirow{3}{*}{100~TeV}
&   MSTW2008 & $33.9^{+7.06\%+0.94\%}_{-8.29\%-1.26\%}$	&
   $57.9^{+8.93\%+0.90\%}_{-9.46\%-1.20\%}$	&
   $0.585^{+1.29\%+0.0526\%}_{-2.02\%-0.0758\%}$ \\ 
&   CT10 & $32.4^{+6.87\%+2.29\%}_{-8.11\%-2.95\%}$ &
   $55.5^{+8.73\%+2.16\%}_{-9.27\%-2.78\%}$ &
   $0.584^{+1.27\%+0.189\%}_{-1.99\%-0.260\%}$ \\ 
&  NNPDF2.3 & $33.2^{+6.62\%+0.78\%}_{-6.47\%-0.78\%}$	&
  $56.9^{+7.62\%+0.75\%}_{-7.29\%-0.75\%}$	&
  $0.584^{+1.29\%+0.0493\%}_{-2.01\%-0.0493\%}$\\ \hline
\end{tabular}
\end{center}
\caption{Results with NLO QCD corrections at
  13~TeV and 100~TeV, using three different sets of PDF. Results are
  presented together with the renormalization/factorization scale and
  PDF uncertainties. Contrary to Table~\ref{tab:scale}, the $\alpha_S$
  systematics is not included here.}
\label{tab:pdf_tot}
\end{table}
%--------------------------------------------------

%--------------------------------------------------
\begin{table}[b!]
 \begin{center}
 \renewcommand{\arraystretch}{1.3}
 \begin{tabular}{r|c|ccc}
\hline
\rule{0pt}{3.6ex} && $\sigma(t\bar{t}H)$[pb] & $\sigma(t\bar{t}Z)$[pb] & $\frac{\displaystyle \sigma(t\bar{t}H)}{\displaystyle \sigma(t\bar{t}Z)}$ \\
\hline
\multirow{5}{*}{13~TeV}
&   default & $0.475^{+5.79\%}_{-9.04\%}$ &
   $0.785^{+9.81\%}_{-11.2\%}$ &
   $0.606^{+2.45\%}_{-3.66\%}$ \\ 
&   $\mu_0=m_t+m_{H,Z}/2$ & $0.529^{+5.96\%}_{-9.42\%}$ &
   $0.885^{+9.93\%}_{-11.6\%}$ &
   $0.597^{+2.45\%}_{-3.61\%}$ \\ 
&   $m_t=\ytop v=174.1~\gev$ &
   $0.474^{+5.74\%}_{-9.01\%}$	&
   $0.773^{+9.76\%}_{-11.2\%}$	&
   $0.614^{+2.45\%}_{-3.66\%}$\\
&   $m_t=\ytop v=172.5~\gev$ &
   $0.475^{+5.81\%}_{-9.05\%}$	&
   $0.795^{+9.82\%}_{-11.2\%}$	&
   $0.597^{+2.45\%}_{-3.65\%}$\\
&   $m_H=126.0~\gev$ & $0.464^{+5.80\%}_{-9.04\%}$
   &  $0.785^{+9.81\%}_{-11.2\%}$	&
   $0.593^{+2.42\%}_{-3.62\%}$\\ \hline
\multirow{5}{*}{100~TeV}
&   default & $33.9^{+7.06\%}_{-8.29\%}$	&
   $57.9^{+8.93\%}_{-9.46\%}$	&
   $0.585^{+1.29\%}_{-2.02\%}$ \\ 
&   $\mu_0=m_t+m_{H,Z}/2$ & $39.0^{+9.76\%}_{-9.57\%}$ &
   $67.2^{+10.9\%}_{-10.6\%}$ &
   $0.580^{+1.16\%}_{-1.80\%}$ \\ 
&   $m_t=\ytop v=174.1~\gev$ &
   $33.9^{+7.01\%}_{-8.27\%}$	&
   $57.2^{+8.90\%}_{-9.42\%}$	&
   $0.592^{+1.27\%}_{-2.00\%}$\\ 
&   $m_t=\ytop v=172.5~\gev$ &
   $33.7^{+6.99\%}_{-8.31\%}$	&
   $58.6^{+8.93\%}_{-9.46\%}$	&
   $0.576^{+1.27\%}_{-1.99\%}$\\ 
&   $m_H=126.0~\gev$ & $33.2^{+7.04\%}_{-8.28\%}$
   &  $57.9^{+8.93\%}_{-9.46\%}$	&
   $0.575^{+1.25\%}_{-1.95\%}$\\ \hline
\end{tabular}
\end{center}
\caption{Results with NLO QCD corrections at 13~TeV
  and 100~TeV by varying some parameter values. Results are presented
  together with the renormalization/factorization scale
  uncertainties.}
\label{tab:tot}
\end{table}
%--------------------------------------------------

We start by discussing the results at the LO in the EW effects.
The scale variation is performed over the standard range
$0.5\mu_0 \le \mu_{R,F} \le 2\mu_0$, with $\mu_R$ and $\mu_F$ varying
independently. Both scale and PDF
choices are correlated between numerator and denominator when
taking the ratios.  
The resulting scale and MSTW~2008NLO PDF +$\as$ uncertainties, 
for the total cross
sections of the individual processes and of for the ratio, are shown
in Table~\ref{tab:scale}. Notice that the scale uncertainty of the
individual processes, in the range of $\pm 7-10\%$, is reduced to
$\pm 1.5\%$ ($\pm 3\%$) for the ratios at 100 (13) TeV. The PDF
variation is reduced by a factor close to 10, to the few permille level.

To corroborate the great stability of the ratios, we also consider
different PDF sets, showing in Table~\ref{tab:pdf_tot} the results
obtained using the following LHAPDF~5.9.1~\cite{Whalley:2005nh} sets:
MSTW2008 NLO~\cite{Martin:2009iq}, CT10 NLO~\cite{Lai:2010vv} and
NNPDF2.3 NLO~\cite{Ball:2012cx} (in this case, we only consider the
PDF variation, and not the $\as$ systematics).  While the overall
envelope of the predictions for the individual rates includes a $\pm
5\%$ range, the ratio uncertainty due to the PDFs remains at the few
permille level.\bigskip

We explore further variations in our default parameter set 
in Table~\ref{tab:tot}. There, 
we remove the PDF uncertainties, which are
practically unaffected by these parameter changes. 
Choosing the fixed value $\mu_0=m_t+m_{H,Z}/2$ for the central 
choice of the renormalization and factorization scales, modifies the ratio
$\sigma(t\bar{t}H)/\sigma(t\bar{t}Z)$ by $1\%-1.5\%$, consistent with
the range established using the dynamical scale. 

For $m_t$, we consider a variation in the range of
$m_t=173.3\pm0.8$~GeV. We notice that $\sigma(\tth)$ is practically
constant. This is due to the anti-correlation between the increase
(decrease) in rate due to pure phase-space, and the decrease
(increase) in the strength of $\ytop$, when the top mass is lower
(higher). 
 The $\ttz$ process is vice versa directly sensitive to $m_t$
at the level of $\pm 1.5\%$ over the $\pm 0.8$~GeV range, and this
sensitivity is reflected in the variation of the cross-section
ratio. We notice, however, that if we kept the value of $\ytop$ fixed
when we change $m_t$, the dynamical effect on the rate would be
totally correlated, and the ratio would remain constant to within a
few permille, as shown in Table~\ref{tab:tot_yt}. This shows that the ratio
is only sensitive to the strength of $\ytop$, and only minimally to
the precise value of $m_t$.

Finally, we observe a $\sim 2\%$ shift in $\sigma(\tth)$ (and
therefore in the ratios) when $m_H$ is
changed by 1~GeV, which is a gross underestimate of the precision with which the
Higgs mass is~\cite{Aad:2015zhl} and will soon be known.\bigskip

%--------------------------------------------------
\begin{table}[t]
 \begin{center}
 \renewcommand{\arraystretch}{1.3}
 \begin{tabular}{r|c|ccc}
\hline
\rule{0pt}{3.6ex} && $\sigma(t\bar{t}H)$[pb] & $\sigma(t\bar{t}Z)$[pb] & $\frac{\displaystyle \sigma(t\bar{t}H)}{\displaystyle \sigma(t\bar{t}Z)}$ \\
\hline
\multirow{2}{*}{13~TeV}
&   $m_t=174.1~\gev$ & $0.3640$	&  $0.5307$ & $0.6860$\\ 
&   $m_t=172.5~\gev$ & $0.3707$	&  $0.5454$ & $0.6800$\\ 
\hline
\multirow{2}{*}{100~TeV}
&   $m_t=174.1~\gev$ & $23.88$	&  $37.99$ & $0.629$\\ 
&   $m_t=172.5~\gev$ & $24.21$	&  $38.73$ & $0.625$\\ \hline
\end{tabular}
\end{center}
\caption{LO results at at~TeV and 100~TeV, keeping
  the top Yukawa coupling $\ytop v =173.3$ GeV.}
\label{tab:tot_yt}
\end{table}
%--------------------------------------------------

%--------------------------------------------------
\setlength{\tabcolsep}{3pt}
\begin{table}[b!]
% \begin{center}
 \renewcommand{\arraystretch}{1.3}
 \begin{tabular}{@{}r|c|ccc|ccc}
\hline
& & \multicolumn{3}{c|}{$\alpha(m_Z)$ scheme} &
   \multicolumn{3}{c}{$G_\mu$ scheme} \\ 
\rule{0pt}{3.6ex}  && $\sigma(t\bar{t}H)$[pb] & $\sigma(t\bar{t}Z)$[pb] &
   $\frac{\displaystyle \sigma(t\bar{t}H)}{\displaystyle \sigma(t\bar{t}Z)}$ & 
   $\sigma(t\bar{t}H)$[pb] & $\sigma(t\bar{t}Z)$[pb] &
   $\frac{\displaystyle \sigma(t\bar{t}H)}{\displaystyle \sigma(t\bar{t}Z)}$ \\ 
\hline
\multirow{5}{*}{13~TeV} 
& NLO QCD 
   & $0.475$ & $0.785$ & $0.606$ 
   & $0.462$ & $0.763$ & $0.606$ \\
& $\mathcal{O}(\alpha_S^2\alpha^2)$ Weak 
   & $-0.006773$ &  $-0.02516$ &  
   & $0.004587$ &  $-0.007904$ & \\ % $-0.5803$
& $\mathcal{O}(\alpha_S^2\alpha^2)$ EW 
   & $-0.0045$ & $-0.022$ & ~ 
   & $0.0071$ & $-0.0033$ & ~ \\
& NLO QCD+Weak 
   & $0.468$ & $0.760$ & $0.617$ 
   & $0.467$ & $0.755$ & $0.619$ \\ 
& NLO QCD+EW 
   & $0.471$ & $0.763$ & $0.617$ 
   & $0.469$ & $0.760$ & $0.618$ \\ \hline
\multirow{5}{*}{100~TeV} 
& NLO QCD 
   & $33.9$ & $57.9$ & $0.585$ 
   & $32.9$ & $56.3$ & $0.585$\\
& $\mathcal{O}(\alpha_S^2\alpha^2)$ Weak 
   & $-0.7295$   & $-2.146$    & 
   & $0.0269$   & $-0.8973$    & \\ % $-0.0300$ 
& $\mathcal{O}(\alpha_S^2\alpha^2)$ EW 
   & $-0.65$ & $-2.0$ & ~ 
   & $0.14$ & $-0.77$ & ~  \\ 
& NLO QCD+Weak 
   & $33.1$ & $55.8$ & $0.594$
   & $32.9$ & $55.4$ & $0.594$\\
& NLO QCD+EW 
   & $33.2$ & $55.9$ & $0.594$
   & $33.1$ & $55.6$ & $0.595$\\ \hline
\end{tabular}
%\end{center}
\caption{Effect of the EW NLO corrections, in the
  $\alpha(m_Z)$ and $G_{\mu}$ schemes, at 13~TeV and 100~TeV.}
\label{tab:default}
\end{table}
\setlength{\tabcolsep}{6pt}
%--------------------------------------------------

The effect of the NLO EW corrections in the  $\alpha(m_Z)$ scheme is shown in
Table~\ref{tab:default}. The shift in
the ratio with respect to the pure NLO QCD result 
is of the order of $2\%$. For reference, we also provide 
the results in
the $G_{\mu}$ scheme. In this scheme, we use
$\alpha^{-1}=132.50699632834286$ and $G_{\mu}=1.166390\cdot
10^{-5}$. The overall difference from the $\alpha(m_Z)$ scheme
for the individual rates is at the
percent level, and at the permille level for the ratios. We conclude
that, once the known NLO EW effects are incorporated, the
residual uncertainty of the cross-section ratio 
due to higher-order EW corrections should be significantly
below the percent level.\bigskip

Before closing this discussion of the total rates, we remark on the
relation between the predictions for the cross section ratios at LO
and at NLO. Since at LO the renormalization and factorization scales
only appear in the PDFs and in $\alpha_{\sss S}(\mu_R)$, and given
that the numerical values of the scales is very similar (as a result
of $m_H-m_Z \ll 2m_t+m_{H,Z}$), the LO ratios come with an unreliably
optimistic estimate of the scale and PDF uncertainty. It is only at
NLO that, through the introduction of the appropriate kinematical
factors $Q^2$ in the renormalization logarithms $\log(\mu_R^2/Q^2)$,
the relevant differences between the scale behavior of two processes
are first exposed. At NLO one also encounters classes of (IR- and
UV-finite) diagrams that differ between the two processes, and
contribute to finite NLO terms that cannot be estimated using
scale-variation arguments. For example, light-quark loops can couple
to the $Z$ boson, but not to the Higgs.
 
As always when using scale-variation tests to assess the theoretical
systematics, there is therefore no guarantee that yet higher-order
corrections will not exceed the range predicted by those
estimates. For a measurement as important as the extraction of the top
Yukawa coupling, it is reasonable to demand that the uncertainty
estimates we provided here be confirmed by a full NNLO calculation,
something that will certainly be possible over the next few
years. Nevertheless we believe that the studies presented here provide
a rather compelling case to argue that a precision at the percent
level is reasonable. 

%%%%%%%%%%%%%%%%%%%%%%%%%%%%%%%%%%%%%%%%%%%%%%%%%%%
\subsection{Kinematical distributions}

Any experimental analysis, and in particular the boosted approach that
we employ in this work, will restrict the phase-space available to the
final states. To preserve the precision in the theoretical prediction
of the ratio of total $\tth$ and $\ttz$ cross sections, it is crucial
to ensure that the reduction in systematics uncertainties carries over
to the description of final states after kinematical cuts have been
applied. We present here a summary of our studies at 100 TeV, focused
on the kinematical distributions most relevant for our studies, and
limited to main sources of uncertainty (scale and PDF). The results
for other distributions and for other systematics (top mass, EW
scheme), at 100 and at 13~TeV, lead to similar results, and are
available upon request.

%--------------------------------------------------
\begin{figure}[t]
 \centering
 \includegraphics[width=0.32\textwidth]{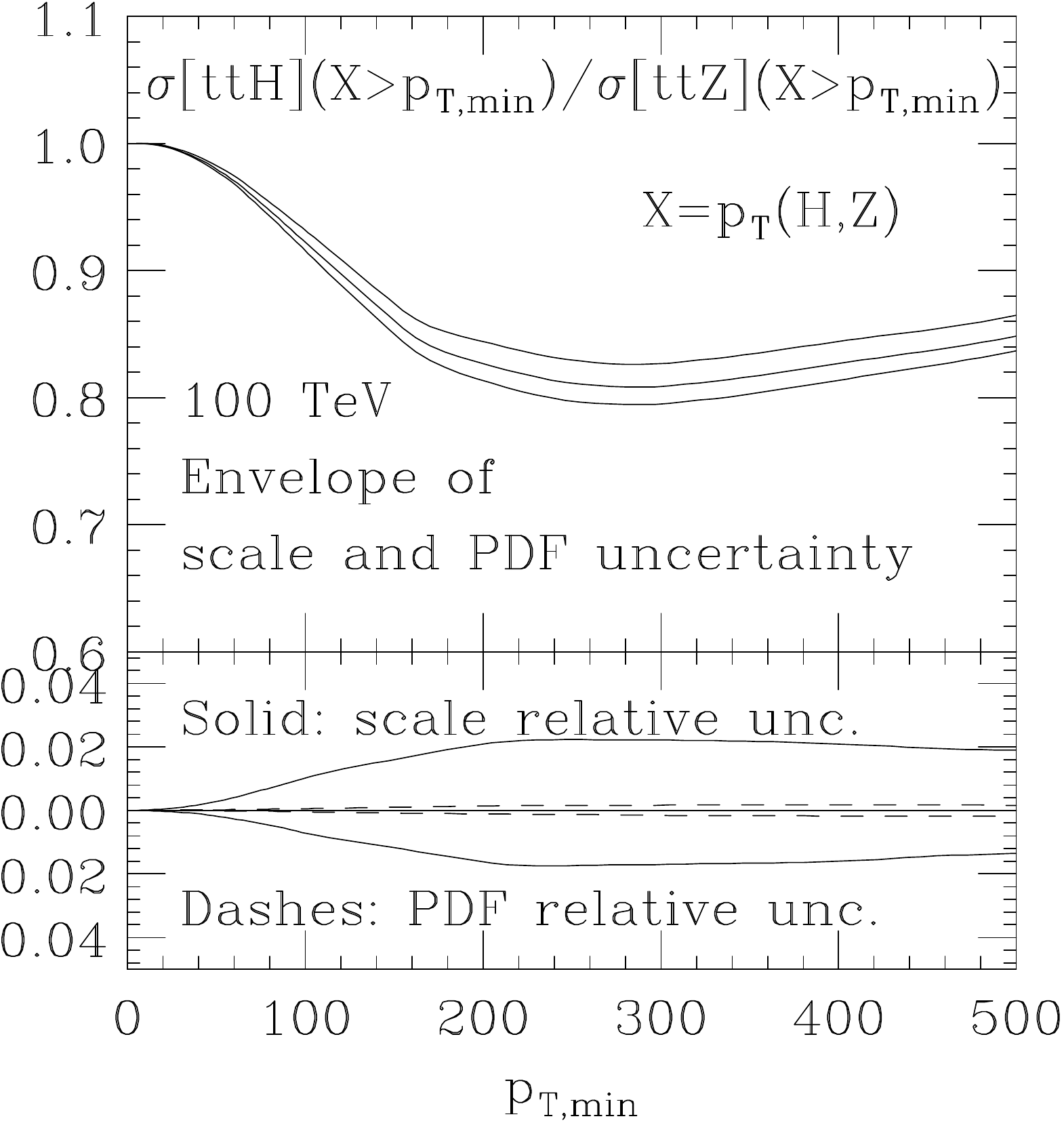}
 \includegraphics[width=0.32\textwidth]{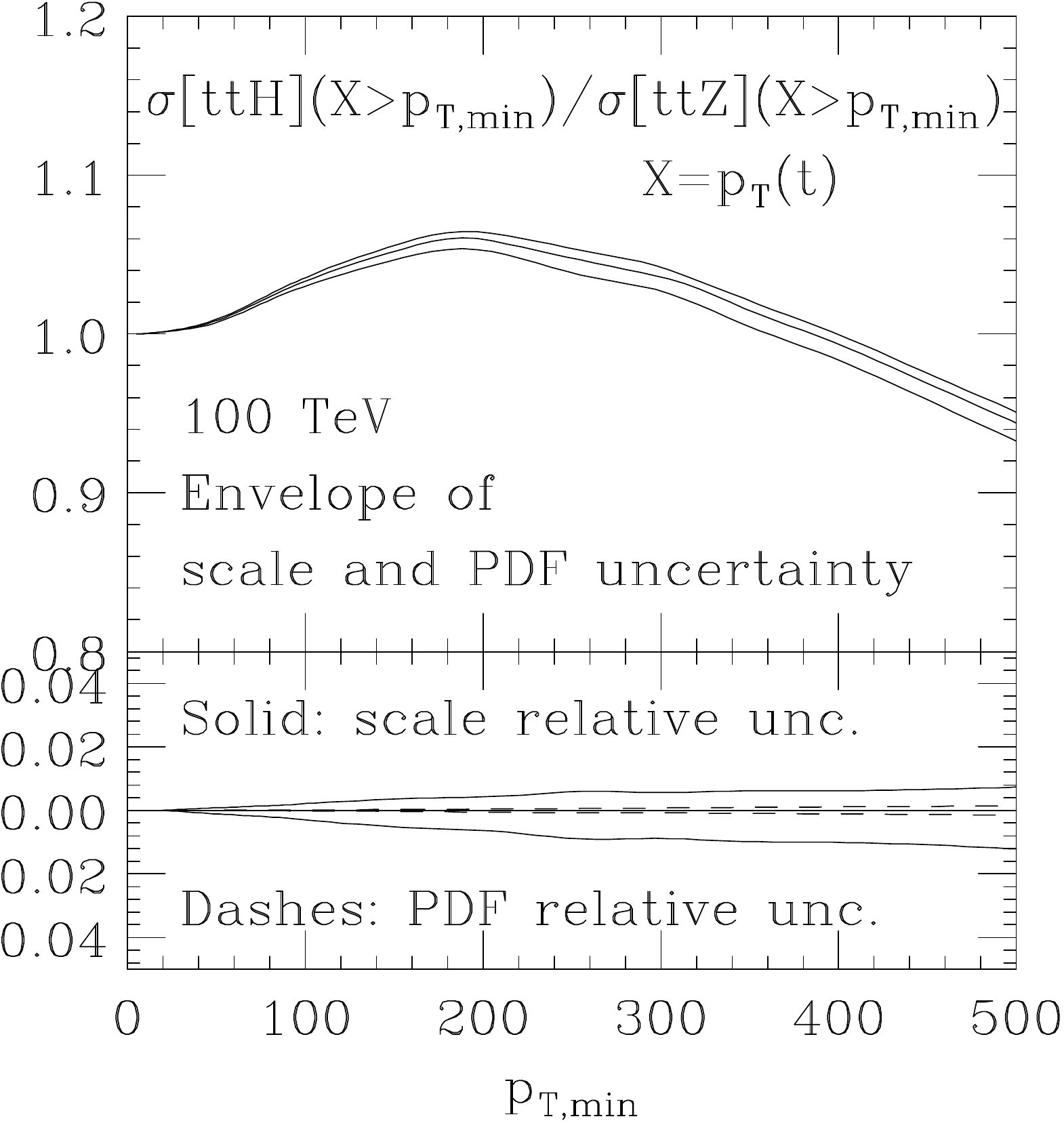}
 \includegraphics[width=0.32\textwidth]{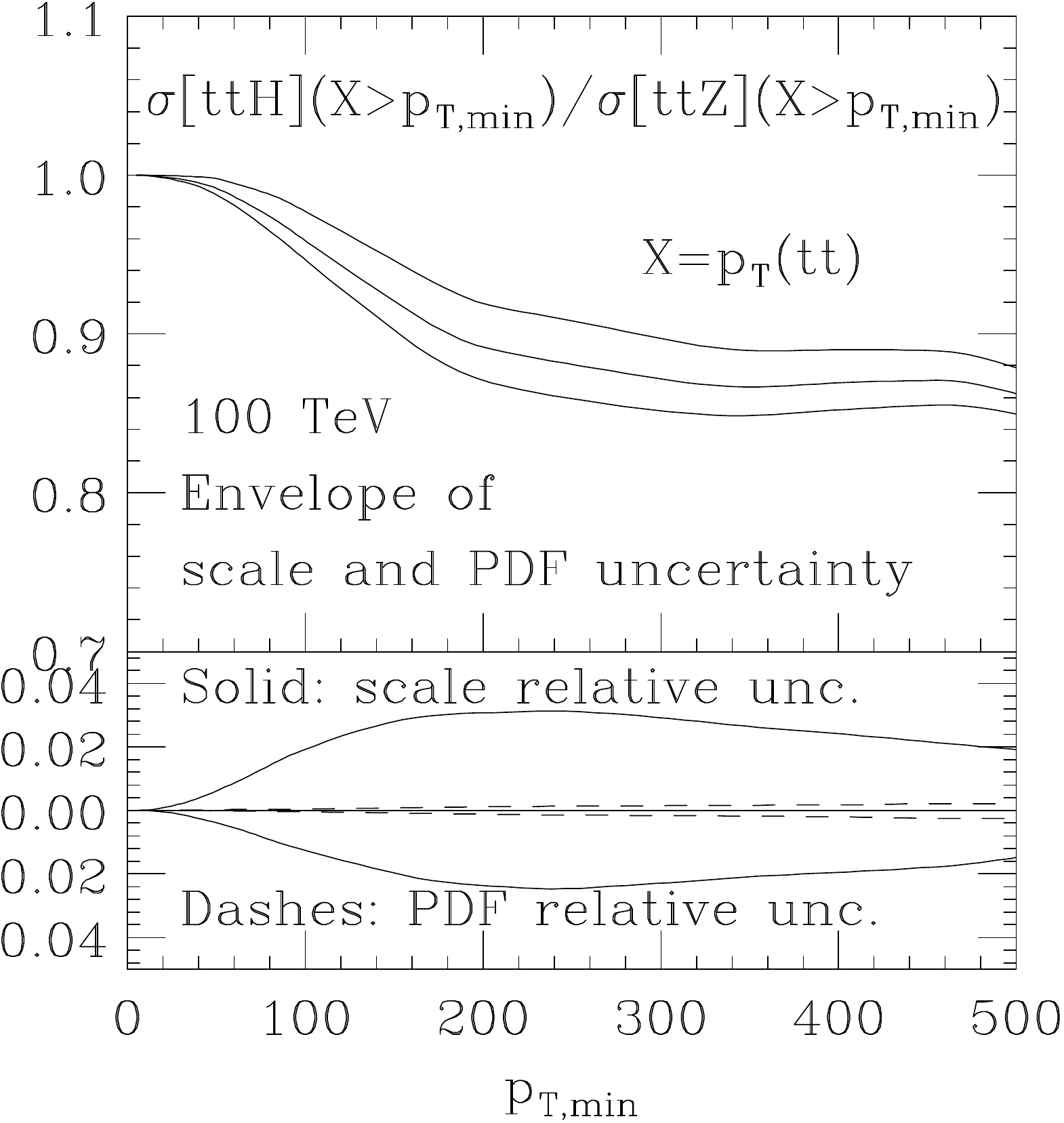}
 \hspace*{0.05\textwidth}
\caption{Scale and PDF systematics of ratios of integrated $p_T$
  spectra for different observables, at 100~TeV. From left to right:
  $p_T$ of the boson, $p_T$ of the top quark, $p_T$ of the $t\bar{t}$
  pair. }
\label{fig:ptsyst}
\end{figure}
%--------------------------------------------------

We show in Fig.~\ref{fig:ptsyst} the ratio of the integrated $p_T$
spectra of various final-state objects $X$:
$\sigma[\tth](p_{T,X}>p_{T,{\rm min}})/ \sigma[\ttz](p_{T,X}>p_{T,{\rm
    min}})$.  On the left, $X=H(Z)$ for the $\tth$ $(\ttz)$
process. In the middle, $X=t$ and on the right $X$ is the $t\bar{t}$
system.  We normalize the ratios to 1 at $p_{T,{\rm min}}=0$, so that
the resulting uncertainties correspond to the systematics in the
extrapolation of the ratio of differential distributions to the ratio
of the total rates.  The three upper panels show that the ratios are
not a constant, and can change buy up to 20\% up to $p_T = 500$~GeV.
The relative uncertainties, separately for the scale and PDF variation
(MSTW2008 NLO set), are shown in the lower plots.  The scale
uncertainties reach a value of $\pm 2\%$ for the boson $p_T$ spectra,
$\pm 1\%$ for the top, and $\pm 3\%$ for the $p_T$ of the $t\bar{t}$
pair.  The PDF uncertainties remain well below the percent level
throughout.

These results imply that the relative shapes of the $p_T$ spectra  
can be controlled with a precision that remains consistent with
the overall goal of a percent-level extraction of the relative
rates. There is no doubt that future NNLO calculations of both
processes will improve this even further.
Very precise measurements of the shape of the $Z$ boson spectra in
$\ttz$ events using e.g. the very clean leptonic $Z$ decay will also
help confirming the accuracy of the predicted $p_T$ spectra
and reduce a possible left-over uncertainty. 

%%%%%%%%%%%%%%%%%%%%%%%%%%%%%%%%%%%%%%%%%%%%%%%%%%%
\section{Boosted $\bf t\bar{t}H$ at 100 TeV}
\label{sec:default}

Just like at the LHC, the $\tth$ production process can be studied for
a variety of Higgs decay channels.  We collect in
Table~\ref{tab:tthdecays} the event rates for potentially interesting
Higgs decays combined with $\tth$ production, for an integrated
luminosity of $20~\iab$ at 100~TeV. These numbers include the
branching ratio for the mixed lepton-hadron $t\bar{t} \to \ell
\nu_\ell +$~jets decay ($\ell=e,\mu$), in addition to the relevant
Higgs branching ratios.

%--------------------------------------------------
\begin{table}[b!]
 \begin{center}
\begin{tabular}{c|c|c|c} 
\hline
$H\to 4\ell$ & $H\to \gamma \gamma$ & $ H\to 2\ell 2\nu$ & $H\to
  b\bar{b}$ \\ \hline
$2.6 \cdot 10^4$ & $4.6 \cdot 10^5$ & $2.0 \cdot 10^6$ & $1.2 \cdot 10^8$ \\ \hline
\end{tabular}
\end{center}
\caption{$\tth$ event rates for various Higgs decay modes, with $20~\iab$ at
  100~TeV, assuming $t\bar{t} \to \ell \nu+$jets. Here and for Higgs
  decays, $\ell$ can be either an electron or a muon.} 
\label{tab:tthdecays}
\end{table}
%--------------------------------------------------

Considering that analysis cuts and efficiencies will typically reduce
these rates by a further factor of 10 or more, it is clear that the
otherwise very clean $H\to 4\ell$ does not have the minimum number of
$10^4$ events, required to aim for a 1\% target precision.  In the
case of $H\to \gamma\gamma$ (see also~\cite{fabio}), we considered a
simple parton-level analysis, implementing basic cuts such as:
\begin{eqnarray}
p_{T, \gamma,b,j}>25~\gev \; &,& \qquad \vert \eta_{\gamma,b,j} \vert <2.5 
\; , \qquad \Delta R_{jj,bb,bj} > 0.4
\nn \\
p_{T,\ell}>20~\gev \; &,& \qquad \vert \eta_\ell \vert <2.5
\end{eqnarray}
These leave around $5 \cdot 10^4$ events with $20~\iab$, while the
$t\bar{t}\gamma\gamma$ background, subject to a $\vert
m_{\gamma\gamma}-125 \vert < 5~\gev$ cut, is almost a factor of 10
smaller. On the other hand, detection efficiencies, such as those
related to lepton or photon isolation and to $b$ tagging, make this
channel borderline for a 1\% statistical accuracy, and call for a
dedicated study including realistic projections of detector
performance.  The $H\to 2\ell 2\nu$ final state has a potentially
interesting rate, which may deserve a separate study.\bigskip

Given the extraordinary rate for the $H\to b\bar{b}$ final state, and
following the original LHC analysis~\cite{heptop_1}, we focus on this
channel,
\begin{equation}
pp \to t\bar{t}H \to (b j j)\, (\bar{b} \ell \bar{\nu})\, (b\bar{b}),
(b \ell \nu)\, (\bar{b} j j)\, (b\bar{b}) \; . 
\end{equation}
The leptonic
top decay guarantees the triggering and reduces multi-jet
combinatorics. The leading backgrounds are:
\begin{itemize}
\item[] $pp \to t \bar t \, b \bar b$, the main irreducible QCD background

\item[] $pp \to t \bar t Z$, including the $Z$-peak in the $m_{bb}$
  distribution 
\item[] $pp \to t \bar t +$jets with fake-bottoms tags 
\end{itemize}
Additional backgrounds like $W$+jets will be small and do not
lead to dangerous kinematical features for our
analysis~\cite{heptop_1}.  The analysis strategy based on boosted top
and Higgs decays is extremely simple~\cite{heptop_1},
\begin{enumerate}
\item an isolated lepton
\item a tagged top without any $b$-tag requirement
\item a tagged Higgs with two $b$-tags inside
\item a continuum $b$-tag outside the top and Higgs fat jets
\end{enumerate}
The $m_{bb}$ distribution will provide us with simple sidebands to
control the $t\bar{t}b\bar{b}$ and $t\bar{t}$+jets backgrounds, and a
second mass peak from the $t\bar{t}Z$ mass peak. We discuss the
unfortunate need for the continuum $b$-tag below. The simplicity of
our analysis will allow us to efficiently control systematics.\bigskip

For simplicity, all Monte Carlo event samples are generated at
leading order. The main effects from the available higher order
predictions of the $t\bar{t}H$ signal~\cite{tth_nlo,tth_dec}, the $t\bar{t}Z$
background~\cite{ttz_nlo}, the $t\bar{t}b\bar{b}$
background~\cite{ttbb_nlo}, and the $t\bar{t}$+jets
background~\cite{ttjj_nlo} are discussed separately in Section~\ref{sec:theory},
so for the signal--background analysis we leave them out.
We use \textsc{MadGraph5}~\cite{madgraph} with
\textsc{NNPDF23} parton densities~\cite{pdf}, showering and
hadronization via \textsc{Pythia8}~\cite{pythia8} and the fast
detector simulation with
\textsc{Delphes3}~\cite{delphes,detector_card}. 

At the generator level we require $p_{T,j,b,\ell}> 10$~GeV and $\Delta
R_{jj,bb,j\ell}> 0.1$.  The $t\bar{t}+$jets background is generated as
one hard jet with $p_{T,j} > 100$~GeV at the hard matrix element
level. We do not consider merged samples since we found that the
influence of $t\bar{t}+2j$ to our analysis is negligible. After generator
cuts we start with a signal cross section of 4.2~pb. Associated
$t\bar{t}Z$ production yields 1.2~pb. The continuum $t\bar{t}b\bar{b}$
background counts 121~pb and is at this stage dominated by
$t\bar{t}+\mathrm{jets}$ with 2750~pb.

\textsc{Delphes3} provides isolated leptons as well as
parton-level $b$-quarks needed for the tagging procedure
later-on. Leptons have to pass a minimum $p_{T,\ell}>10$~GeV. For
their isolation we demand a transverse momentum ratio (isolation
variable) of $I < 0.1$ within $\Delta R<0.3$.  Finally, we use the
energy flow objects for hadrons to cluster via the Cambridge/Aachen
(C/A) jet algorithm~\cite{ca_algo}. The jet clustering and the
analysis are done with \textsc{FastJet3}~\cite{fastjet}, a modified
BDRS Higgs tagger~\cite{bdrs,heptop_1} and the
\textsc{HEPTopTagger2}~\cite{heptop_new}. For all $b$-tags we require a
parton-level $b$-quark within $\Delta R < 0.3$. \bigskip

%--------------------------------------------------
\begin{figure}[t]
 \centering
 \includegraphics[width=0.45\textwidth]{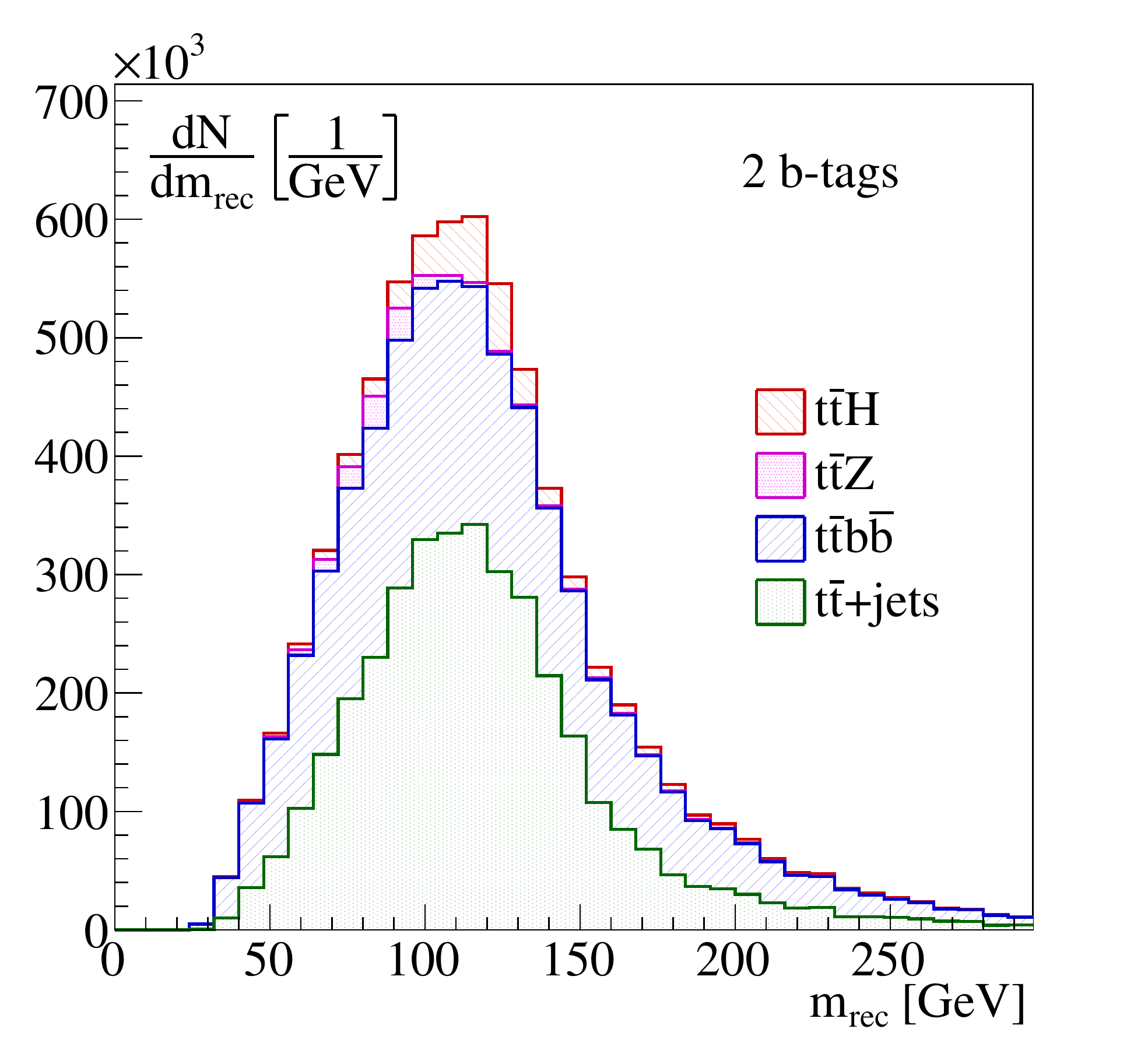}
 \hspace*{0.05\textwidth}
 \includegraphics[width=0.45\textwidth]{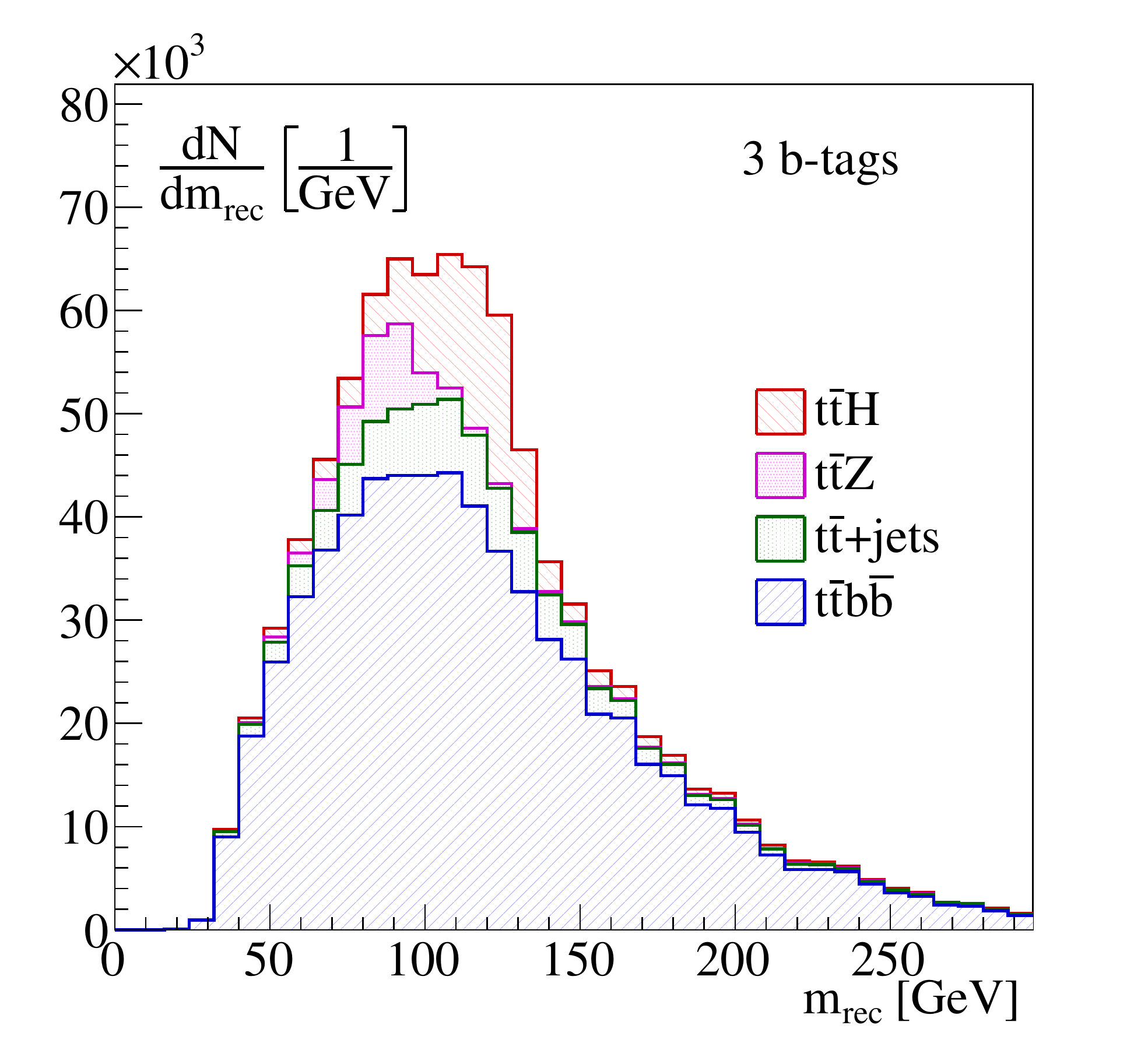}
 \caption{Reconstructed $m_{bb}$ for the leading-$J$ substructures in
   the fat Higgs jet.  We require two $b$-tags inside the fat Higgs
   jet (left) and an additional continuum $b$-tag (right).  The event
   numbers are scaled to $\mathcal{L} = 20~\iab$.}
\label{fig:btags}
\end{figure}
%--------------------------------------------------

First, we require one isolated lepton with $|y_\ell|<2.5$ and
$p_{T,\ell}>15$~GeV. For the top
tag~\cite{tag_history,heptop_2,tag_review}, we cluster the event into
fat C/A jets with $R=1.8$ and $p_{T,j}>200$~GeV. Provided we find at
least two fat jets we apply the \textsc{HEPTopTagger2} with the
kinematic requirement $|y_{j}^{(t)}|<4$.  The recent significant
update of the \textsc{HEPTopTagger2} relies on two additional pieces
of information to achieve a significant
improvement~\cite{heptop_new}. One of them is
$N$-subjettiness~\cite{nsubjettiness}, which adds some sensitivity to
the color structure of the event.  The other is the optimalR mode,
which based on a constant fat jet mass reduces the size of the fat
jet~\cite{variableR} to the point where the fat jet stops containing
all hard top decay subjets. This minimal size can also be computed
based on the transverse momentum of the fat jet. Since the signal and
all considered backgrounds include a hadronic top quark, changing the
top tagging parameters results only in an overall scaling factor. In
this analysis we do not cut on the difference between the expected and
the found optimal radius because the initial fat jet size is already
chosen to fit the expected transverse momenta. To have a handle on the
QCD multi-jet background, we place a mild cut on the filtered
$N$-subjettiness ratio $\tau_3/\tau_2 < 0.8$ which can be tightened at
the cost of signal efficiency if desired.  After identifying the
boosted top we remove the associated hadronic activity and apply a
modified BDRS Higgs tagger to fat C/A jet(s) with $R=1.2$,
$|y_j^{(H)}|<2.5$, and $p_{T,j}>200$~GeV. Our decomposition of the fat
jet into hard substructure includes a cutoff of
$m_\mathrm{sub} > 40$~GeV for the relevant substructure and a mass
drop threshold of 0.9. The hard substructures are then paired in all
possible ways and ordered by their modified Jade distance,
\begin{equation}
J=p_{T,1}p_{T,2}(\Delta R_{12})^4 \; .
\end{equation}
The leading pairing we filter~\cite{bdrs} including the three hardest
substructures, to allow for hard gluon radiation. For consistency we
require a reconstructed transverse momentum above 200~GeV. Within this
Higgs candidate we ask for two $b$-tags, assuming a global tagging
efficiency of $50\%$ and a mis-tagging probability of $1\%$ for all
jets within $|y_j| < 2.5$ and $p_{T,j} > 30$~GeV. As we can see in the
left panel of Fig.~\ref{fig:btags}, the $t\bar{t}$+jets and
$t\bar{t}b\bar{b}$ backgrounds are of similar size at this
stage. Moreover, the analysis sculpts the backgrounds towards
$m_{bb} \sim 100$~GeV.\bigskip

To simplify the background composition and to avoid the strong
background sculpting it turns out that a third, continuum $b$-tag is
useful. We target the decay jet of the otherwise leptonically decaying
top by removing the top and Higgs constituents from the event and then
clustering the remaining hadronic structure into C/A jets with $R=0.6$
and $p_{T,j}>30$~GeV. For one of them we require a $b$-tag within
$|y_b| < 2.5$ and a angular separation $\Delta R_{b,j}>0.4$ from all
other jets, now including the top and Higgs decay products.

The effect of this third $b$-tag becomes clear in the right panel of
Fig.~\ref{fig:btags}. We are now dominated by the continuum
$t\bar{t}b\bar{b}$ background. The corresponding event rates for an
integrated luminosity of $20~\iab$ are given in Tab.~\ref{tab:rates}.
While the light-flavor $t\bar{t}$+jets background is now suppressed
well below the leading $t\bar{t}b\bar{b}$ background it is still of
the same size as the Higgs signal, which means we still need to
include it in our analysis.

%--------------------------------------------------
\begin{table}[t!]
\begin{center} \begin{small}
\begin{tabular}{l|rrl} 
\hline
$m_{bb}\in [100, 150]$~GeV & 2 $b$-tags & 3 $b$-tags & ratio\\
  \hline
$t \bar t H$        & 2.4E+5 & 6.4E+4 & 1/3.8 \\
$t \bar t b \bar b$ & 1.2E+6 & 2.4E+5 & 1/5.0 \\
$t \bar t$+ jets    & 1.9E+6 & 3.8E+4 & 1/50  \\
$t \bar t Z$        & 2.3E+4 & 4.9E+3 & 1/4.7 \\
\hline
\end{tabular}
\end{small} \end{center}
\caption{Event rates assuming an integrated luminosity of $20~\iab$.}
\label{tab:rates}
\end{table}
%--------------------------------------------------

%%%%%%%%%%%%%%%%%%%%%%%%%%%%%%%%%%%%%%%%%%%%%%%%%%%
\section{Updated BDRS tagger}
\label{sec:update}

\begin{figure}[b!]
\includegraphics[width=0.246\textwidth]{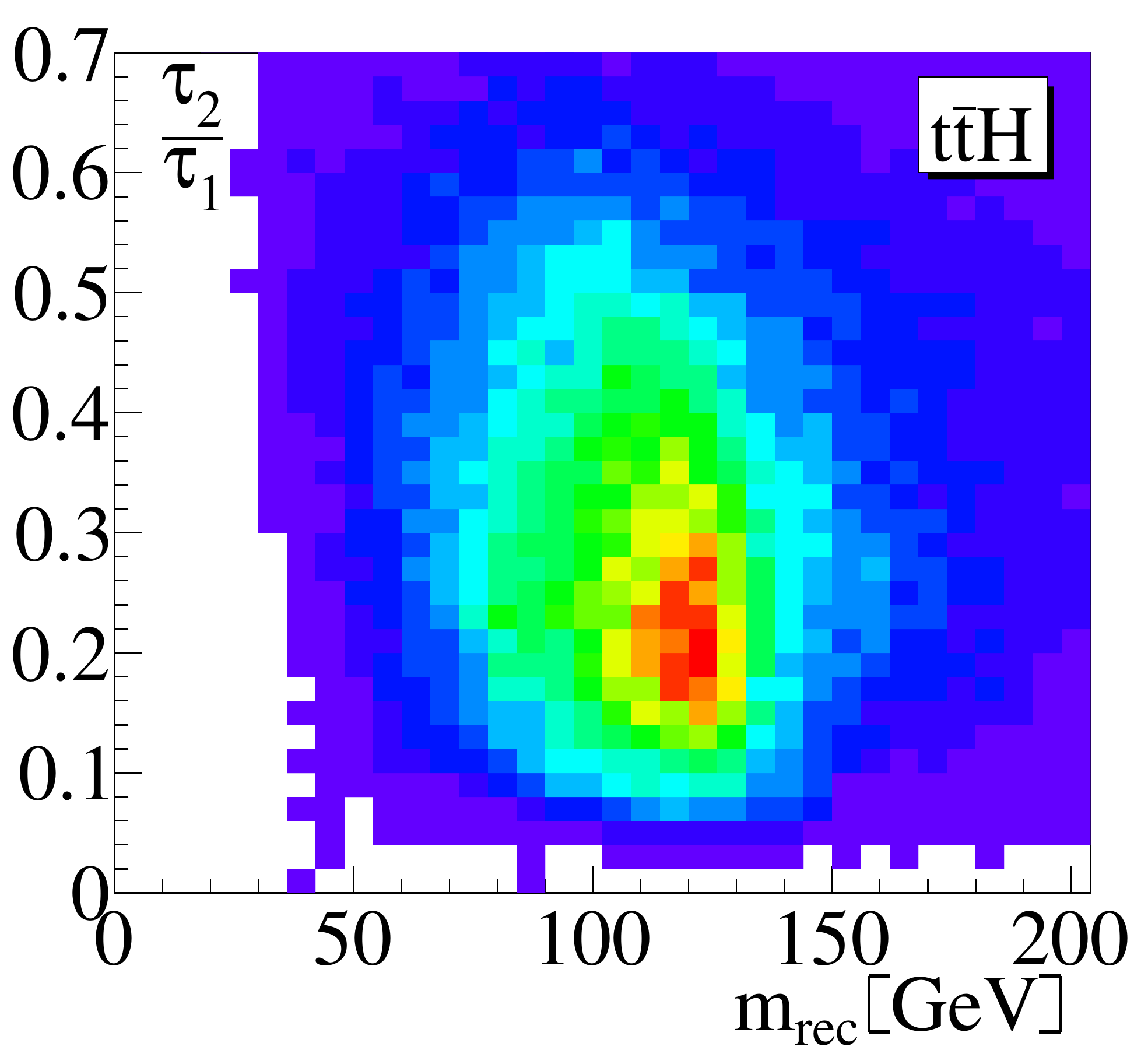}
\includegraphics[width=0.246\textwidth]{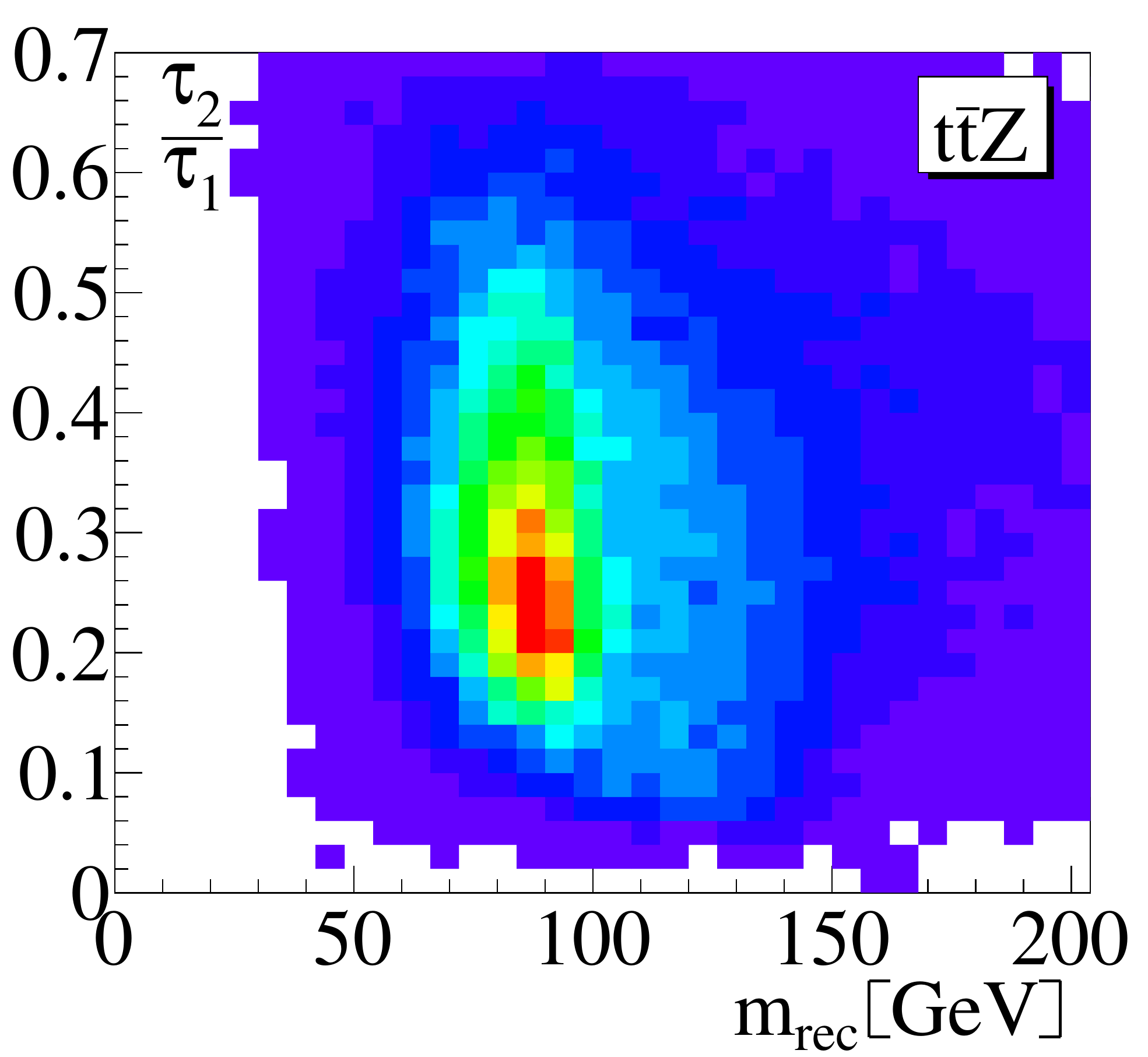}
\includegraphics[width=0.246\textwidth]{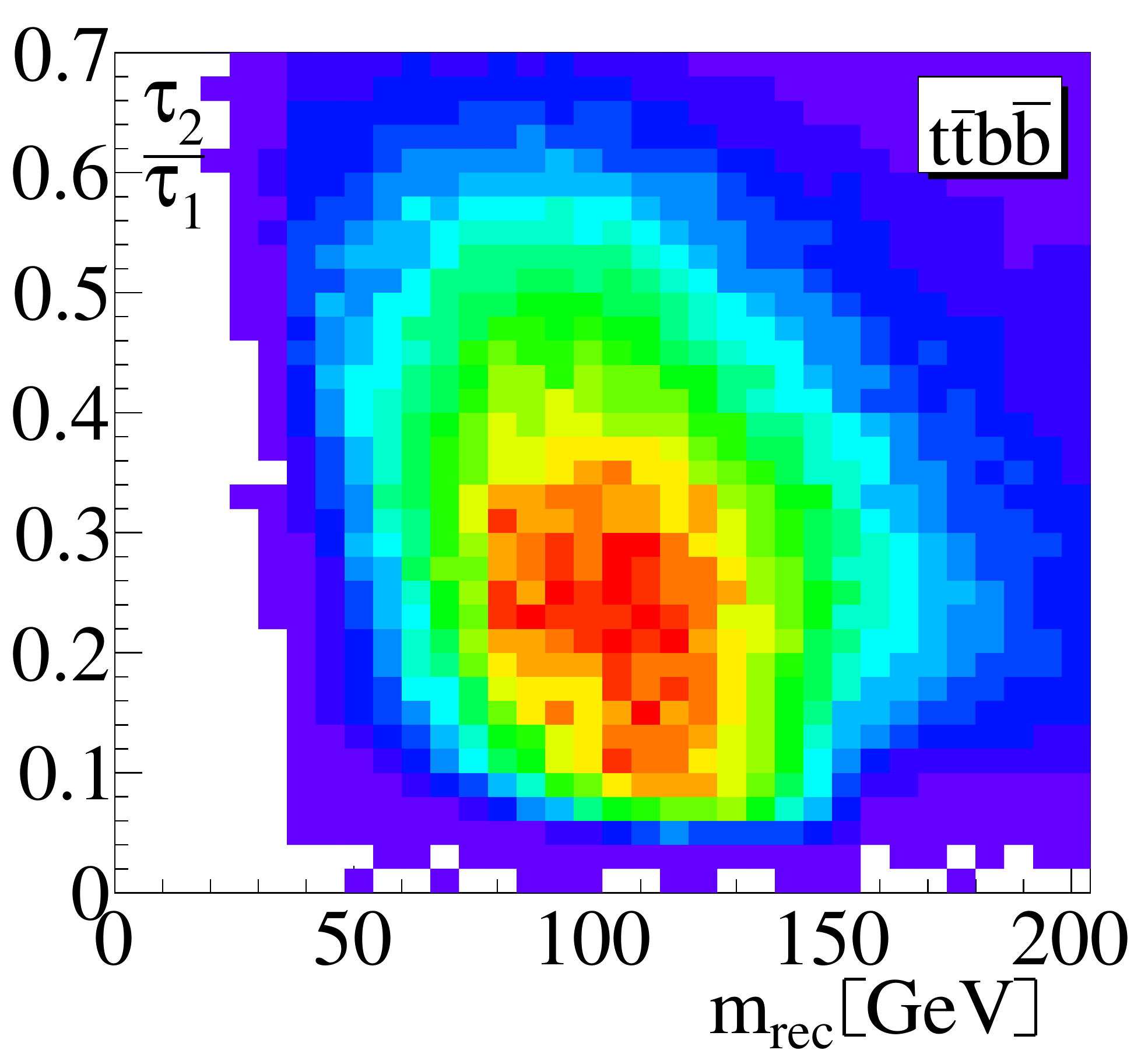}
\includegraphics[width=0.246\textwidth]{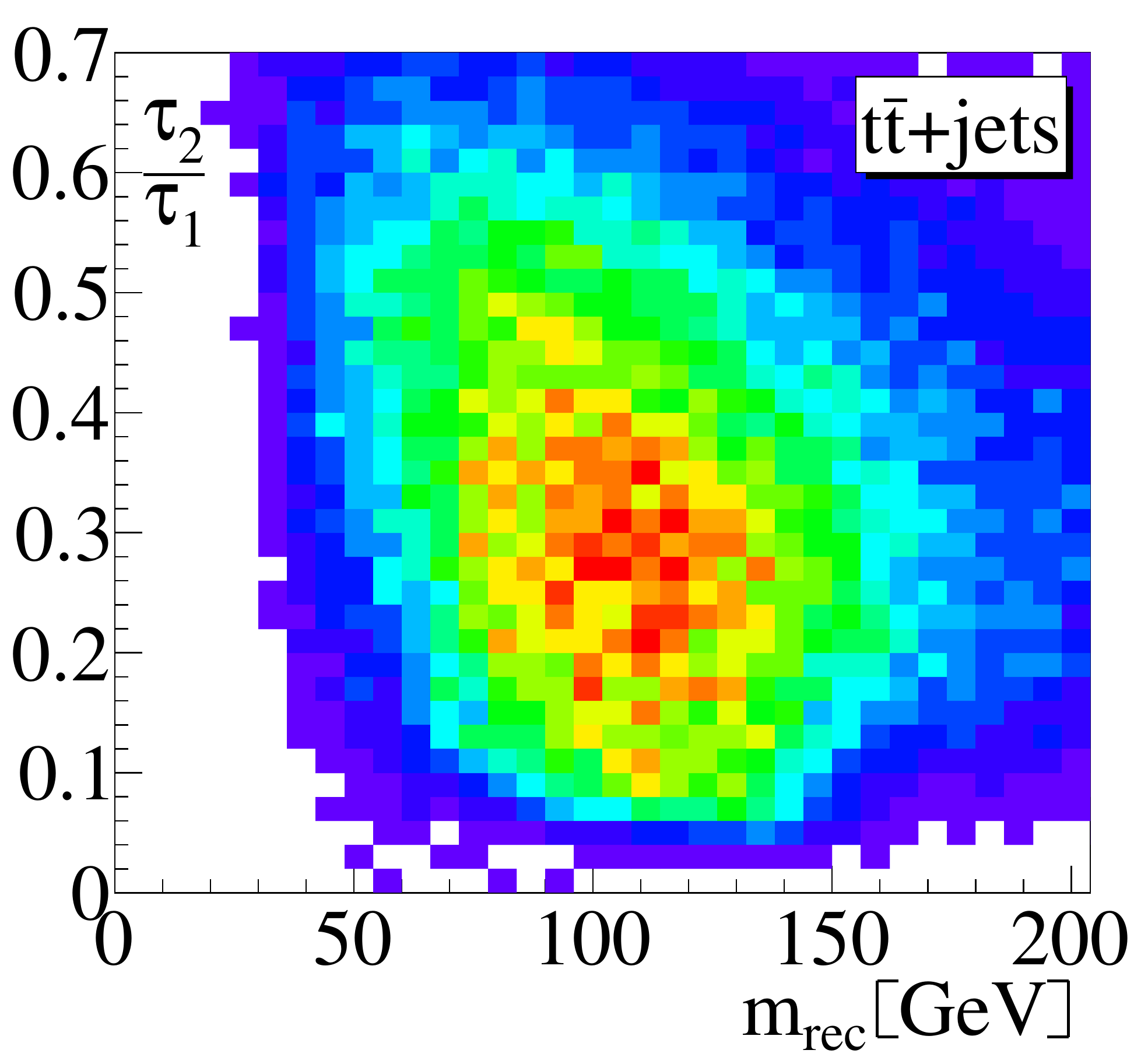}
\caption{Correlation between the reconstructed mass $m_\mathrm{rec}$
  and the $N$-subjettiness ratio $\tau_2/\tau_1$ of the filtered Higgs
  candidate fat jet for the signal and background samples. The event
  numbers are scaled to $\mathcal{L} = 20~\iab$.}
\label{fig:nsubj_cut} 
\end{figure}

\begin{figure}[t!]
 \centering
 \includegraphics[width=0.45\textwidth]{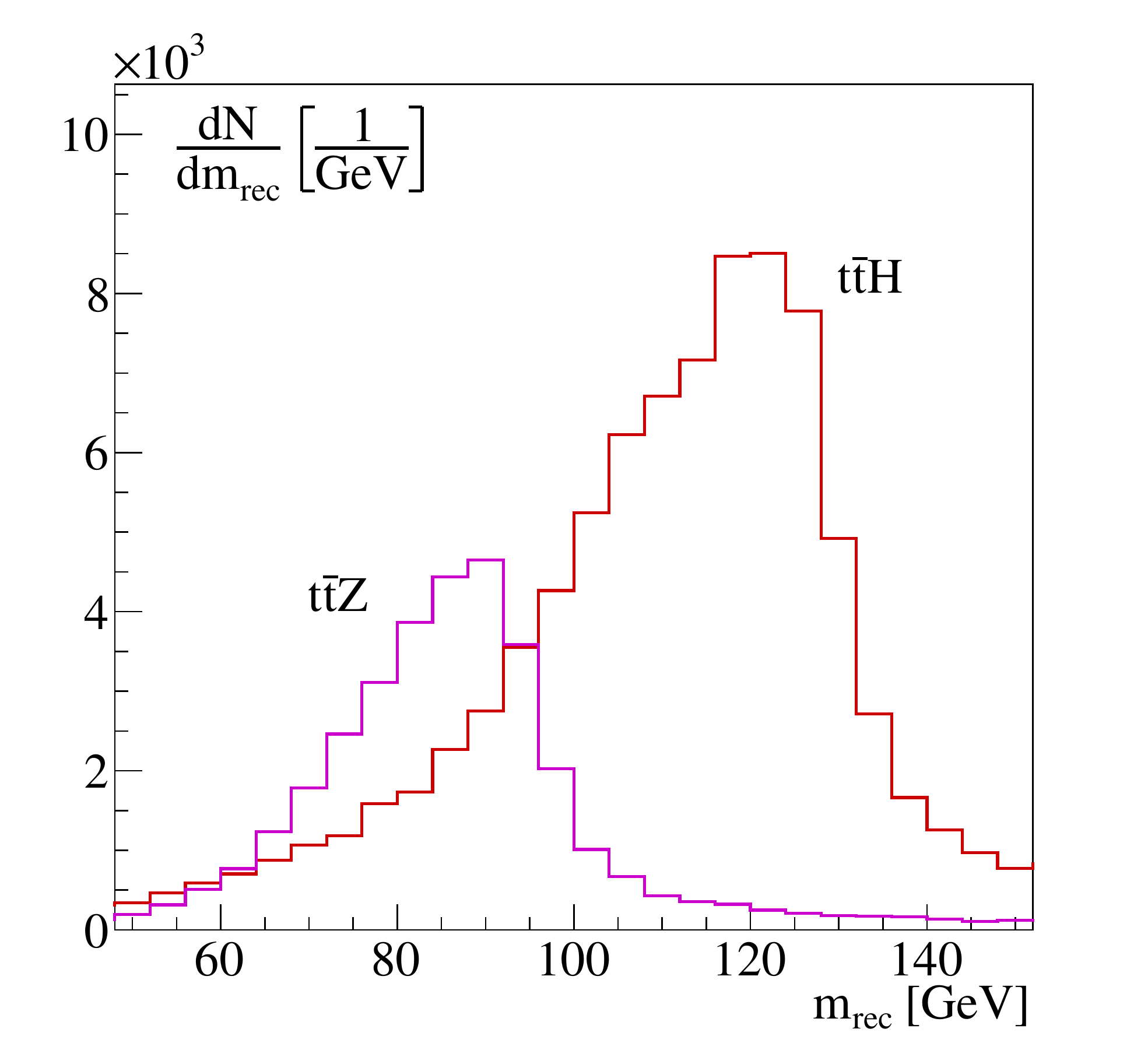}
 \hspace*{0.05\textwidth}
\includegraphics[width=0.45\textwidth]{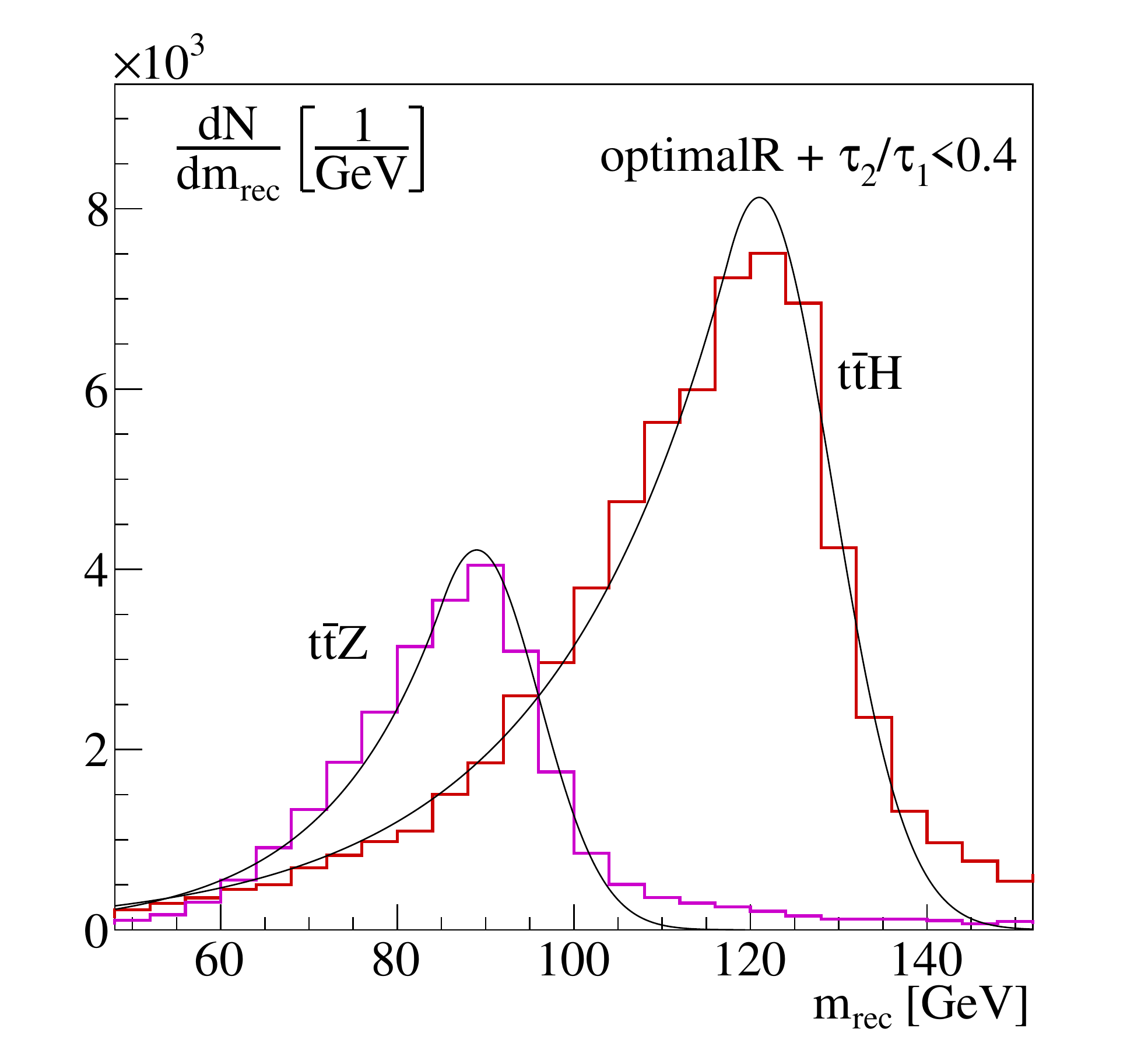}
\caption{Reconstructed $m_{bb}$ of the Higgs and $Z$ candidates in $t
  \bar t H$ and $t\bar{t}Z$ production with the default BDRS tagger
  (left) and after using optimalR and the $N$-subjettiness cut
  $\tau_2/\tau_1<0.4$ (right). In the right panel we include the
  fitted Crystal Ball functions. The event numbers are scaled to
  $\mathcal{L} = 20~\iab$.}
\label{fig:nsubj_HZpeaks}
\end{figure}

The two improvements of the \textsc{HepTopTagger2} can also be added
to the BDRS Higgs tagger~\cite{bdrs,fastjet,bdrs_ours}. The decay $H
\to b \bar b$ will typically contain two hard substructures, so using
$N$-subjettiness the characteristic parameter $\tau_2/\tau_1$ has to
be small.  The correlations between the reconstructed masses and the
ratio $\tau_2/\tau_1$ of the filtered fat jets in
Fig.~\ref{fig:nsubj_cut} indicate that a cut $\tau_2/\tau_1<0.4$ not
only reduces the backgrounds but additionally leads to narrower and
better-defined mass peaks for the Higgs and $Z$-decays as shown in
Fig.~\ref{fig:nsubj_HZpeaks}. \bigskip

%--------------------------------------------------
%--------------------------------------------------

%--------------------------------------------------
%--------------------------------------------------

\begin{figure}[!t]
\centering
\includegraphics[width=0.45\textwidth]{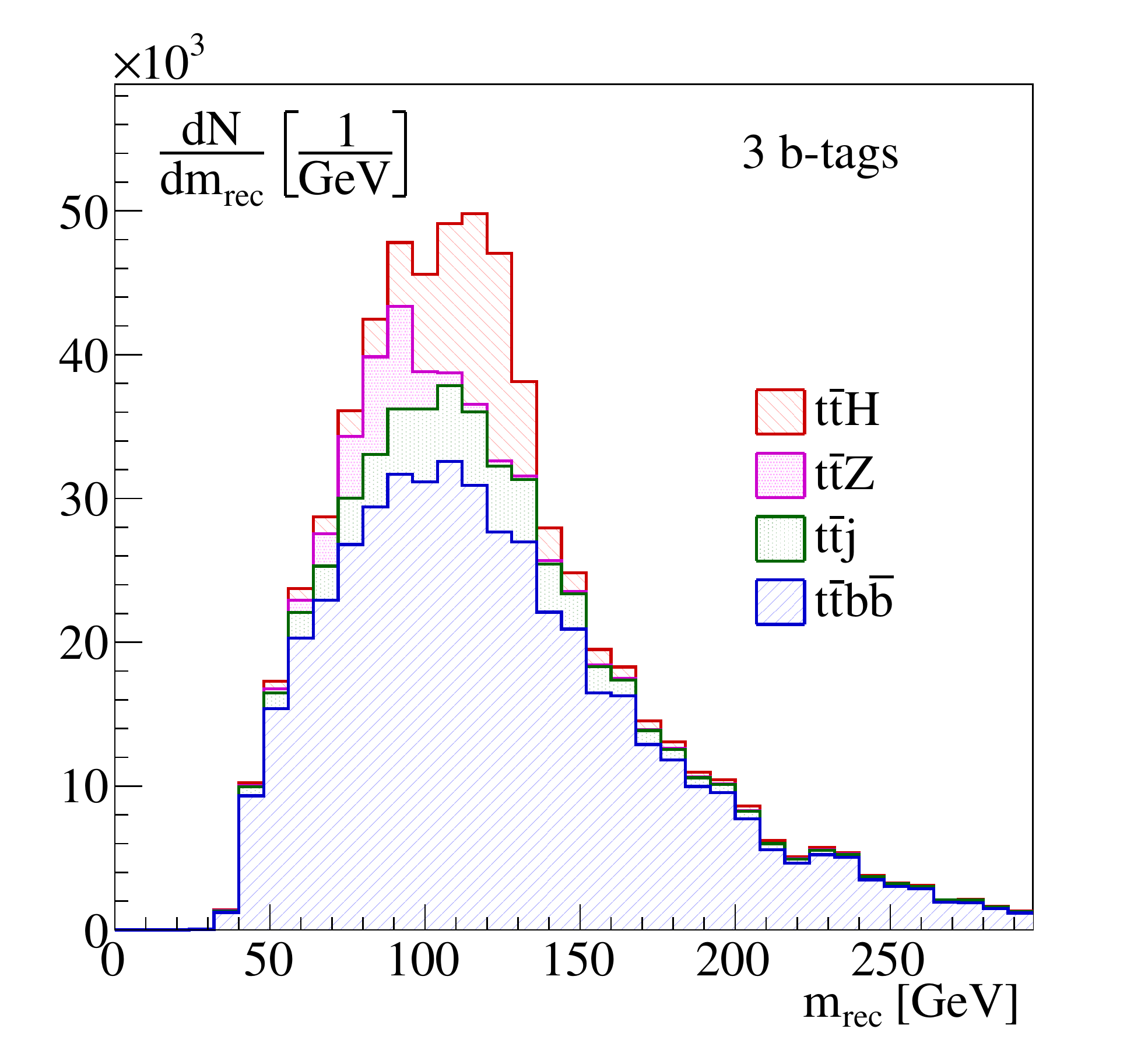}
 \hspace*{0.05\textwidth}
\includegraphics[width=0.45\textwidth]{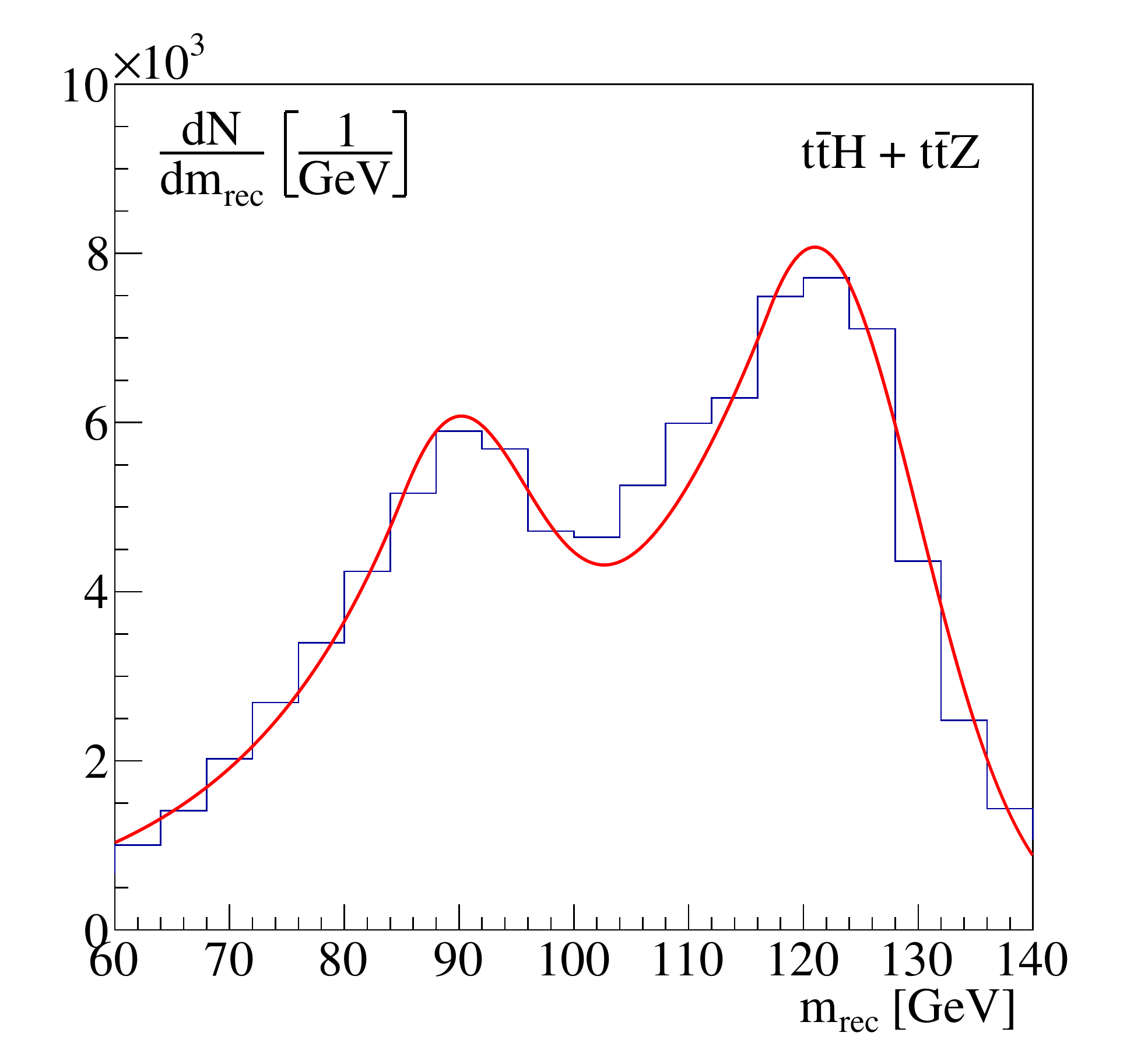}
 \caption{Left: Reconstructed $m_{bb}$ for the leading-$J$
   substructures in the fat Higgs jet. We require two $b$-tags inside
   the fat Higgs jet and a continuum $b$-tag. Unlike in
   Fig.~\ref{fig:btags} we apply an $N$-subjettiness cut and use an
   optimalR version of the BDRS tagger. Right: Double-peak fit
   assuming perfect continuum background subtraction. The event
   numbers are scaled to $\mathcal{L} = 20~\iab$.}
\label{fig:nsubj} 
\end{figure}

In the optimalR version of the BDRS tagger we reduce the size of the
Higgs fat jet candidate. Aside from reduced underlying event and
pile-up this minimizes the combinatorics in the $m_{bb}$
reconstructions. As for the top case~\cite{heptop_new} we shrink the
fat jet radius in steps of 0.1 as long as the jet mass does not drop
below $m_j <0.8 \, m_{j,\mathrm{orig}}$ relative to the originally
tagged Higgs jet with $R=1.2$. We can extract the expected value of
$\Delta R_{bb}$ from a fit to Monte Carlo simulations,
\begin{equation}
\centering
\Delta R_{bb}^\mathrm{(calc)} =
\frac{250~\gev}{p_{T,\mathrm{filt}}} \; .
\label{eq:Rfunc}
\end{equation}
This supports the choice of $R=1.2$ for the C/A jet clustering for the
Higgs Tagger requiring transverse momenta of
$p_T>200$~GeV. Unfortunately, for $t\bar{t}H$ production the relation
between the expected and the measured values of $\Delta R_{bb}$ does
not significantly improve the analysis. However, the mass difference
between the Higgs and the $Z$ boson leads to a shifted peak in the
$\Delta R_{bb} - \Delta R_{bb}^\mathrm{(calc)}$ distribution for
$\ttz$. This shift allows for an additional reduction of $\ttz$ if
desired. In the final result shown in the left panel of
Fig.~\ref{fig:nsubj} we include a triple $b$-tag, the $N$-subjettiness
variable $\tau_2/\tau_1$, and a modified fat jet radius for the Higgs
candidate. Since the background region $m_\mathrm{bb} \in
[160,300]$~GeV is smooth and untouched by any signal, we can use it to
subtract the QCD continuum from the combined $t \bar t H$ and $t \bar
t Z$ signal. If the soft regime $m_\mathrm{bb} \in [0,60]$~GeV can be
useful in the same way needs to be checked by a full experimental
analysis.\bigskip

For the signal region $m_{bb} \in [104,136]$~GeV we arrive at a
signal-to-background ratio around $S/B \approx 1/3$ and a Gaussian
significance $S/\sqrt{B}=120$, assuming an integrated luminosity of
$\mathcal{L} = 20~\iab$. The error on the number of nominally
$N_S = 44700$ signal events is given by two terms. First, we assume
that we can determine $N_S$ from the total number of events $N_S+N_B$
using a perfect determination of $N_B$ from the side bands. Second,
the side band $m_{bb} \in [160,296]$~GeV with altogether
$N_\mathrm{side} = 135000$ events and a relative uncertainty of
$1/\sqrt{N_\mathrm{side}}$ introduces a statistical uncertainty
$\Delta N_B$, altogether leading to
\begin{eqnarray}
\Delta N_S 
&=& 
\left[ \left( \sqrt{N_S + N_B} \right)^2 
      +\left( \Delta N_B \right)^2
\right]^{1/2} \nonumber \\
&=& 
\left[ \left( \sqrt{N_S + N_B} \right)^2 
      +\left( \frac{N_B}{\sqrt{N_\mathrm{side}}} \right)^2
\right]^{1/2} = 0.013 \, N_S \; .
\end{eqnarray}
For the Yukawa coupling this translates into a relative error of
around 1\%. The first term alone would give
$\Delta N_S = 0.010 \, N_S$. \bigskip

In the right panel of Fig.~\ref{fig:nsubj} we show a combined fit to
the $Z$ and Higgs peaks assuming a perfect background subtraction. A
combined analysis of both peaks (with known masses) serves as a check
of the jet substructure techniques~\cite{bdrs,heptop_1} and as a means
to reduce systematic and theoretical uncertainties, as discussed in
Section~\ref{sec:theory}.  Given separate simulations
for the Higgs and $Z$ peaks, we can fix the shape of both distributions
by fitting a Crystal Ball function~\cite{crystal_ball} to each of them
as done in the right panel of Fig.~\ref{fig:nsubj_HZpeaks}. For these
fits we limit the exponent of the non-Gaussian tails to 50. In
addition we fix the peak positions accounting for a shift due to
losses in the reconstruction. Their linear combination allows us to
model the background subtracted mass distribution. In the combined fit
we keep all shape parameters fixed and allow only for separate scaling factors
of each peak. From the double Crystal Ball function we finally receive
the relative size of the two peak areas ${N_H}/{N_Z} = 2.80 \pm 0.03$.
Using the combined fit therefore allows us to probe the top Yukawa
coupling with a statistical precision of $\sim 0.5\%$. 
Given the discussion of Section~\ref{sec:theory}, this precision can
be eventually matched by the theoretical systematics, assuming no new
physics affects $t\bar{t}Z$ production beyond the percent level. It
remains to be explored to which extent the future detectors can
benefit from the potential cancellations of experimental systematics
in the measurement of the ${N_H}/{N_Z}$ ratio.

%%%%%%%%%%%%%%%%%%%%%%%%%%%%%%%%%%%%%%%%%%%%%%%%%%%
\section{Outlook}

%--------------------------------------------------
%--------------------------------------------------

The top Yukawa coupling is one of two key parameter required for the
understanding of the Higgs potential, and it is a crucial ingredient
to the measurement of the Higgs self-coupling. At the LHC its
determination will be limited to around $\Delta \ytop /\ytop \approx 10\%$
because of statistical as well as theoretical
uncertainties~\cite{sfitter_future,Brock:2014tja}. 
At a 100~TeV hadron collider the
increased statistics will significantly improve this measurement.

We proposed here to measure the top Yukawa coupling using the
decay $H \to b\bar{b}$ in the boosted phase space regime. Our simple
analysis strategy~\cite{heptop_1} relies on a trigger lepton and two
fat jets, one from the hadronic Higgs decay and one from the hadronic
top decay. The $m_{bb}$ distribution will show a clear peak from the
Higgs signal as well as a similarly large peak from the $Z$
background. The continuum side band and the second peak offer two ways
to control the backgrounds as well as the translation of the $t\bar{t}
\, b\bar{b}$ rate into a measurement of the Yukawa coupling. We
find that a measurement of the top Yukawa coupling to around $1\%$
should be feasible at 100~TeV collider energy with an integrated
luminosity of $20~\iab$. This is an order of magnitude improvement
over the expected LHC reach, with significantly improved
control over the critical uncertainties.

There exist additional, complementary handles on the uncertainties.
For example, the $H\to \gamma\gamma$ decay could allow a direct
measurement of the ratio of branching ratios $B(H\to
\gamma\gamma)/B(H\to b\bar{b})$. It would serve as be complementary,
although indirect, probe of the $\tth$ coupling. Furthermore, $H\to
2\ell 2\nu$ could also be interesting, since there is enough rate to
explore the regime $p_{T,H} \gg m_H$, which, particularly for the $e^\pm
\mu^\mp \nu\bar{\nu}$ final state, could prove particularly
clean.\bigskip

\begin{center} \textbf{Acknowledgments} \end{center}\medskip

TS would like to thank the International Max Planck Research School for
\textsl{Precision Tests of Fundamental Symmetries} for their support.
HSS would like to thank S. Frixione, V. Hirschi, D.
Pagani and M. Zaro for useful discussions and for collaborating on the
EW project in {\sc MadGraph5\_aMC@NLO}.
The work of MLM and HSS is supported by the ERC grant 291377
\textsl{LHCtheory: Theoretical predictions and analyses of LHC physics:
advancing the precision frontier}.

%%%%%%%%%%%%%%%%%%%%%%%%%%%%%%%%%%%%%%%%%%%%%%%%%%%


\begin{thebibliography}{10}
\bibitem{bdrs_ours}
 The corresponding code is available from the authors.
\end{thebibliography}
\end{document}